\documentclass[aps,eqsecnum,nofootinbib,superscriptaddress,showpacs,preprintnumbers]{revtex4}
\usepackage{amssymb, amsmath, amsopn, amsthm,mathptmx}
\usepackage{epsfig,amsfonts,bm}
\usepackage{MnSymbol}

\def\eref#1{Eq.\ \eqref{#1}}
\def\erefs#1{Eqs.\ \eqref{#1}}

\begin{document}
\preprint{KOBE-COSMO-16-01}

\title{Probing circular polarization in stochastic gravitational wave background \\with pulsar timing arrays}
\date{\today}
\begin{abstract}
We study the detectability of circular polarization in a
stochastic gravitational wave background from various sources such as supermassive black hole binaries, cosmic strings, 
and inflation in the early universe
 with pulsar timing arrays.  
We calculate generalized overlap reduction functions for the circularly polarized stochastic gravitational wave
background. We find  that the circular polarization can not be detected 
for an isotropic background. However, there is a chance to observe the circular polarization for an anisotropic gravitational wave background.
We also show how to separate polarized gravitational waves from unpolarized gravitational waves. 
\end{abstract}
\author{Ryo Kato}
\email{153s107s@stu.kobe-u.ac.jp}
\author{ Jiro Soda}
\email{jiro@phys.sci.kobe-u.ac.jp}
\affiliation{Department of Physics, Kobe University, Rokkodai 1-1, Kobe 657-8501, Japan}


\maketitle

\section{Introduction}\label{sec:introduction}
It is believed that the direct detection of gravitational waves (GWs) will bring the era of gravitational wave astronomy.
 The interferometer detectors are now under operation and awaiting the first signal of GWs~\cite{TheLIGOScientific:2014jea,TheVirgo:2014hva,Somiya:2011np}. 
 It is also known that pulsar timing arrays (PTAs) can be used as a detector for GWs \cite{Detweiler:1979wn,Romani(1989),1990ApJ...361..300F}.
These detectors are used to search for very low frequency ($\sim \rm nHz$) gravitational waves, 
where the lower limit of the observable frequencies is determined by the inverse of total observation time $T$. Indeed, the
total observation time has a crucial role in PTAs, because PTAs are most sensitive near the lower edge of observable frequencies \cite{Lentati:2015qwp,Arzoumanian:2015liz}.
Taking into account its sensitivity, the first direct detection of the gravitational waves might be achieved by PTAs.

The main target of PTAs is the stochastic gravitational wave background (SGWB)  generated by a large number of unresolved 
sources with the astrophysical  origin or the cosmological origin in the early universe.
The promising sources are super massive black hole binaries~\cite{Phinney:2001di}, cosmic (super)string~\cite{Vilenkin:1984ib,Kuroyanagi:2012jf}, and inflation~\cite{Maggiore:2000gv,Grishchuk:2005qe}.
Previous studies have assumed that the SGWB is isotropic and unpolarized~\cite{Allen:1997ad}.
These assumptions are reasonable for the primary detection of the SGWB, but the deviation from the isotropy and the polarizations
 should have rich information of sources of gravitational waves.
Recently, the cross-correlation formalism has been generalized to deal with anisotropy in the SGWB \cite{Mingarelli:2013dsa}.
Result of this work enables us to consider arbitrary levels of anisotropy, and a Bayesian approach was performed by using this formalism \cite{Taylor:2013esa}.
On the other hand, for the anisotropy of the SGWB, the cross-correlation formalism has been also developed
 in the case of interferometer detectors~\cite{Seto:2008sr}. As to the polarization, there are works including the ones motivated by the modified gravity~\cite{Chamberlin:2011ev,Gair:2015hra}.
 We can envisage supermassive black hole binaries emit circularly polarized SGWB due to the Chern-Simons term~\cite{Jackiw:2003pm}.
There may also exist cosmological SGWB with circular polarization in the presence of parity violating term in gravity sector~\cite{Satoh:2007gn,Contaldi:2008yz,Takahashi:2009wc,Cook:2011hg,Obata:2014loa}.

In this paper, we investigate  the detectability of circular polarization in the SGWB by PTAs. 
We characterize  SGWB  by the so called Stokes $V$ parameter~\cite{lightman}  and 
calculate generalized overlap reduction functions (ORFs) so that we can probe the circular polarization of the SGWB.
We also discuss a method to separate the intensity ($I$ mode) and circular polarization ($V$ mode) of the SGWB. 

The paper is organized as follows.
In Section \ref{sec:STOKES PARAMETERS FOR A PLANE GRAVITATIONAL WAVE}, we introduce the Stokes parameters for monochromatic plane
 gravitational waves, and clarify the physical meaning of the Stokes parameters $I$ and $V$.
In Section \ref{sec:Formulation}, we formulate the cross-correlation formalism for anisotropic circularly polarized SGWB with PTAs.
The basic framework is essentially a combination of the formalism of \cite{Mingarelli:2013dsa}, and the polarization decomposition formula of the SGWB 
derived in \cite{Seto:2008sr}.
In section \ref{sec:the generalized overlap reduction function for circular polarization}, we calculate the generalized ORFs for the $V$ mode.
The results for $I$ mode are consistent with the previous work~\cite{Mingarelli:2013dsa}.
In section \ref{sec:separation method}, we give a method for separation between  the $I$ mode and $V$ mode of the SGWB.
The final section is devoted to the conclusion.
In Appendixes, we present analytic results for the generalized overlap reduction functions.
In this paper, we will use the gravitational units $c=G=1$.

\section{Stokes Parameters for  Plane Gravitational Waves}\label{sec:STOKES PARAMETERS FOR A PLANE GRAVITATIONAL WAVE}
Let us consider the Stokes parameters for plane waves traveling in the direction $\hat{\bf n}=z$, which can be described by
\begin{eqnarray}
&&h_{xx}(t,z)=-h_{yy}(t,z)={\rm Re}[B_{+}\mathrm{e}^{-iw(t-z)}] \ ,\\
&&h_{xy}(t,z)=h_{yx}(t,z)={\rm Re}[B_{\times}\mathrm{e}^{-iw(t-z)}] \ .
\end{eqnarray}
For an idealized monochromatic plane wave, complex amplitudes $B_{+}$ and $B_{\times}$ are constants.
Polarization of the plane GWs is characterized by the tensor, (see \cite{lightman} and also electromagnetic case \cite{Rybicki,landau})
\begin{eqnarray}
J_{AA'} = B_{A}^{\ast}B_{A'} 
=
\left(
    \begin{array}{cc}
       B_{+}^{\ast}B_{+} &  B_{+}^{\ast}B_{\times}  \\
       B_{\times}^{\ast}B_{+} &  B_{\times}^{\ast}B_{\times} 
    \end{array}
\right)_{AA'}\ ,
\label{jab}
\end{eqnarray} 
where $A,A'$ take $+,\times$. 
Any $2\times2$ Hermitian matrix can be expanded by the Pauli and the unit matrices with real coefficients.
Hence, the $2\times 2$ Hermitian matrix $J_{AA'}$ can be written as 
\begin{eqnarray}
J_{AA'}=\frac{1}{2}(I\sigma_{0,AA'}+Q\sigma_{3,AA'}+U\sigma_{1,AA'}+V\sigma_{2,AA'}) 
=\frac{1}{2}
\left(
    \begin{array}{cc}
       I+ Q &  U-iV  \\
        U+iV  &  I-Q 
    \end{array}
\right)_{AA'}\ ,\label{iquv}
\end{eqnarray}
where
\begin{eqnarray}
\sigma_{0,AA'}=
\left(
    \begin{array}{cc}
       1 &  0 \\
       0  & 1 
    \end{array}
\right)_{AA'}\ ,\quad
\sigma_{1,AA'}=
\left(
    \begin{array}{cc}
       0 &  1 \\
       1  & 0 
    \end{array}
\right)_{AA'}\ ,\quad
\sigma_{2,AA'}=
\left(
    \begin{array}{cc}
       0 &  -i \\
       i  & 0 
    \end{array}
\right)_{AA'}\ ,\quad
\sigma_{3,AA'}=
\left(
    \begin{array}{cc}
       1 &  0 \\
       0  & -1 
    \end{array}
\right)_{AA'}.  \\ \nonumber
\end{eqnarray}
By analogy with electromagnetic cases, $I,Q,U,$ and $V$ are called Stokes parameters. 
Comparing \eref{jab} with \eref{iquv}, we can read off the Stokes parameters as
\begin{eqnarray}
I&=&|B_{+}|^2+|B_{\times}|^2,\label{sti}\\
Q&=&|B_{+}|^2-|B_{\times}|^2,\\
U&=&2{\rm Re}[ B_{+}^{\ast}B_{\times}]= B_{+}^{\ast}B_{\times}+ B_{\times}^{\ast}B_{+},\\
V&=&-2{\rm Im}[ B_{+}^{\ast}B_{\times}]=i( B_{+}^{\ast}B_{\times}- B_{\times}^{\ast}B_{+}).\label{stv}
\end{eqnarray}
Apparently, the real parameter $I$ is the intensity of GWs. In order to reveal the physical meaning of the real parameter $V$,
 we define the circular polarization bases \cite{misner}
\begin{eqnarray}
e_{ij}^{R}=\frac{e_{ij}^{+}+ie_{ij}^{\times}}{\sqrt{2}} \ ,\quad \,e_{ij}^{L}=\frac{e_{ij}^{+}-ie_{ij}^{\times}}{\sqrt{2}} \ .
\end{eqnarray}
From the relation
\begin{eqnarray}
B_{ij}\equiv B_{R}e_{ij}^{R}+B_{L}e_{ij}^{L}=B_{+}e_{ij}^{+}+B_{\times}e_{ij}^{\times}\label{bij} \ ,
\end{eqnarray}
we see
\begin{eqnarray}
B_{R}=\frac{B_{+}-iB_{\times}}{\sqrt{2}},\quad B_{L}=\frac{B_{+}+iB_{\times}}{\sqrt{2}} \ .
\end{eqnarray}
Thus, we can rewrite the Stokes parameters \eqref{sti}-\eqref{stv} as
\begin{eqnarray}
I&=&|B_{R}|^2+|B_{L}|^2,\\
Q&=&B_{R}^{\ast}B_{L}+ B_{L}^{\ast}B_{R},\\
U&=&-i(B_{R}^{\ast}B_{L}-B_{L}^{\ast}B_{R}),\\
V&=&-|B_{R}|^2+|B_{L}|^2.
\end{eqnarray}
From the above expression, we see that the real parameter $V$ characterizes the asymmetry of circular polarization amplitudes.
The other parameters $Q$ and $U$  have additional information about linear polarizations by analogy with the electromagnetic cases.
Alternatively, we can also define the tensor $J'_{CC'}$ in circular polarization bases
\begin{eqnarray}
J'_{CC} = B_{C}^{\ast}B_{C'} 
= \left(
    \begin{array}{cc}
       B_{R}^{\ast}B_{R} &  B_{R}^{\ast}B_{L}  \\
       B_{L}^{\ast}B_{R} &  B_{L}^{\ast}B_{L} 
    \end{array}
\right)_{CC'} 
    = \frac{1}{2}\left(
    \begin{array}{cc}
       I-V & Q+iU  \\
       Q-iU &  I+V 
    \end{array}
\right)_{CC'}\ ,
\end{eqnarray}
where $C,C'=R,L$. Note that the Stokes parameters satisfy a relation
\begin{eqnarray}
I^2=Q^2+U^2+V^2 \ .
\end{eqnarray}

Next, we consider the transformation of the Stokes parameters under rotations around the $\hat{\bf n}=z$ axis.
The  rotation around the $\hat{\bf n}=z$ axis is given by
\begin{eqnarray}
x^{\mu} \rightarrow x'^{\mu}={\Lambda^{\mu}}_{\nu}x^{\nu} \ ,\qquad
\Lambda  =
\left(
    \begin{array}{cccc}
       1 & 0 & 0 & 0 \\
       0 & \cos\psi & -\sin\psi & 0 \\
       0 & \sin\psi & \cos\psi & 0 \\
       0 & 0 & 0 & 1 \\
    \end{array}
\right)\ ,
\end{eqnarray}
where $\psi$ is the angle of the rotation.
The  GWs traveling in the direction $\hat{\bf n}=z$ 
\begin{eqnarray}
 h_{\mu\nu}(x)=B_{\mu\nu}\mathrm{e}^{-ikz}\ ,
\end{eqnarray}
transform as
\begin{eqnarray}
B'_{\mu\nu} = {\Lambda_{\mu}}^{\rho}{\Lambda_{\nu}}^{\lambda}B_{\rho\lambda} \ ,
\end{eqnarray}
where we took the transverse traceless gauge
\begin{eqnarray}
B_{\rho\lambda} =
\left(
    \begin{array}{cccc}
       0 & 0 & 0 & 0 \\
       0 & B_{+} & B_{\times} & 0 \\
       0 & B_{\times} & -B_{+} & 0 \\
       0 & 0 & 0 & 0 \\
    \end{array}
\right)_{\rho\lambda} \ .
\label{Amn}
\end{eqnarray}
After a short calculation, we obtain
\begin{eqnarray}
B'_{+}&=&B_{+}\cos2\psi-B_{\times}\sin2\psi \ ,\label{a+} \\
B'_{\times}&=&B_{+}\sin2\psi+B_{\times}\cos2\psi \ .\label{a*}
\end{eqnarray}
Using \erefs{a+} and \eqref{a*}, the four Stokes parameters (\ref{sti})-(\ref{stv}) transform as
\begin{eqnarray}
I'&=&I\ ,\\
Q'&=&Q\cos4\psi-U\sin4\psi\ ,\\
U'&=&Q\sin4\psi+U\cos4\psi\ ,\\
V'&=&V\ .
\end{eqnarray}
As you can see, the parameters $Q$ and $U$  depend on the rotation angle $\psi$.
This reflects the fact that $Q$ and $U$ parameters characterize linear polarizations.
Note that this transformation is similar to the transformation of electromagnetic case except for  the angle $4\psi$
and can be rewritten as 
\begin{eqnarray}
I'&=&I \ ,\label{henkan1}\\
Q'+iU'&=&e^{4i\psi}(Q+iU) \ ,\\
Q'-iU'&=&e^{-4i\psi}(Q-iU) \ ,\\
V'&=&V \ . \label{henkan2}
\end{eqnarray}

\section{Formulation}\label{sec:Formulation}
In this section, we study anisotropic distribution of SGWB and focus on the detectability of circular polarizations with pulsar timing arrays.
 We combine the analysis  of \cite{Mingarelli:2013dsa} and that of \cite{Seto:2008sr}. 
In Sec.\ref{subsec:The spectral}, we derive the power spectral density for anisotropic circularly polarized SGWB $S_{h}^{AA'}(f,\hat{\bf n})$.
Then we also derive the dimensionless density parameter $\Omega_{\rm gw}(f)$ which is expressed by the frequency spectrum of intensity 
 $I(f)$~\cite{Mingarelli:2013dsa}.
In Sec.\ref{subsec:The signal}, we extend the generalized ORFs to cases with circular polarizations characterized by the parameter $V$.
 For simplicity, we consider specific anisotropic patterns with $l=0,1,2,3$ expressed by the spherical harmonics $Y_{lm}({\bf\hat n})$.

\subsection{Power spectral density $S_{h}^{AA'}(f,\hat{\bf n})$ and density parameter $\Omega_{\rm gw}(f)$}\label{subsec:The spectral}
In the transverse traceless gauge, metric perturbations $h_{ij}(t,{\bf x})$ with a given propagation direction $\hat{\bf n}$ 
can be expanded as \cite{Maggiore-2008}
\begin{eqnarray}
h_{ij}(t,{\bf x})=\sum_{A=+,\times}\int_{-\infty}^{\infty} df  \int_{S^2} d^2 {\bf\hat n}\,\,
\tilde h_{A}(f,{\bf \hat n})\,\,e_{ij}^{A}(\hat{\bf n})\,\,\mathrm{e}^{{-2\pi if}(t-{\bf\hat n}\cdot{\bf x})} \, \,, \label{hij}
\end{eqnarray}
where the Fourier amplitude satisfies $\tilde h_{A}(-f,{\bf \hat n})=\tilde h_{A}^{\ast}(f,{\bf\hat n})$ 
as a consequence of the reality of $h_{ij}(t,{\bf x})$, $d^2 {\bf\hat n}=d\cos\theta d\phi$, $f$ is the frequency of the GWs, $i,j=x,y,z$ are spatial indices, $A=+,\times$ label polarizations.
Note that the Fourier amplitude $\tilde h_{A}(f,{\bf \hat n})$ satisfies the relation 
\begin{eqnarray}
f^2B_{ij}({f,\bf\hat n})=\sum_{A=+,\times}\tilde h_{A}(f,{\bf \hat n})e_{ij}^{A}(\hat{\bf n}) \ ,\label{relation}
\end{eqnarray}
where $B_{ij}({f,\bf\hat n})$ was defined by \eref{bij}.
The polarized tensors $e_{ij}^{A}(\hat{\bf n})$ are defined by
\begin{eqnarray}
e_{ij}^{+}(\hat{\bf n})=\hat {\bf u}_i\hat {\bf u}_j-\hat {\bf v}_i\hat {\bf v}_j \ ,\\
e_{ij}^{\times}(\hat{\bf n})=\hat {\bf u}_i\hat {\bf v}_j+\hat {\bf v}_i\hat {\bf u}_j \ ,
\end{eqnarray}
where $\bf u$ and $\bf v$ are unit orthogonal vectors perpendicular to $\bf\hat n$. 
The polarization tensors satisfy
\begin{eqnarray}
e^A_{ij}(\hat{\bf n})e^{A'\,ij}(\hat{\bf n})=2\delta^{AA'} \ .\label{ee}
\end{eqnarray}
With polar coordinates,  the direction $\hat{\bf n}$ can be represented by
\begin{eqnarray}
\hat {\bf n}=(\sin\theta\cos\phi,\sin\theta\sin\phi,\cos\theta)\ ,\label{kika1}
\end{eqnarray}
and the polarization basis vectors read
\begin{eqnarray}
\hat {\bf u}=(\sin\phi,-\cos\phi,0) \ ,\quad
\hat {\bf v}=(\cos\theta\cos\phi,\cos\theta\sin\phi,-\sin\theta) \ .\label{kika2}
\end{eqnarray}

We assume the Fourier amplitudes $\tilde h_{A}(f,\hat{\bf n})$ are random variables, which is stationary and  Gaussian. 
However, they are not isotropic and unpolarized.
The ensemble average of Fourier amplitudes can be written as \cite{Seto:2008sr,Mingarelli:2013dsa}
\begin{eqnarray}
\langle \tilde h_{A}^{\ast}(f,\hat{\bf n})\tilde h_{A'}(f',\hat{\bf n}')\rangle=\delta(f-f')\delta^2(\hat{\bf n},\hat{\bf n}')S_{h}^{AA'}(f,\hat{\bf n})\ ,\label{pawa-}
\end{eqnarray}
where
\begin{eqnarray}
S_{h}^{AA'}(f,\hat{\bf n})=
\left(
    \begin{array}{cc}
       I(f,\hat{\bf n})+ Q(f,\hat{\bf n}) &  U(f,\hat{\bf n})-iV(f,\hat{\bf n})  \\
        U(f,\hat{\bf n})+iV(f,\hat{\bf n})  &  I(f,\hat{\bf n})- Q(f,\hat{\bf n}) 
    \end{array}
\right) \ .\label{sh}
\end{eqnarray}
Here, the bracket $\langle ...\rangle$ represents an ensemble average, 
and $\delta^2(\hat{\bf n},\hat{\bf n}')$ is the Dirac delta function on the two-sphere.
 The GW power spectral density $S_{h}^{AA'}(f,\hat{\bf n})$ is a Hermitian matrix, and satisfies $S_{h}^{AA'}(f,\hat{\bf n})=S_{h}^{A'A}(-f,\hat{\bf n})$ because of the relation $\tilde h_{A}(-f,{\bf \hat n})=\tilde h_{A}^{\ast}(f,{\bf\hat n})$. Therefore, we have the relations
\begin{eqnarray}
I(f,\hat{\bf n})= I(-f,\hat{\bf n}) \ ,\quad
Q(f,\hat{\bf n})= Q(-f,\hat{\bf n}) \ ,\quad
U(f,\hat{\bf n}) =U(-f,\hat{\bf n}) \ ,\quad
V(f,\hat{\bf n})=-V(-f,\hat{\bf n}) \ .
\end{eqnarray}
Note that the Stokes parameters are not exactly the same as the expression of \eref{iquv}, 
but they have the relation \eref{relation} and characterize the same polarization.
We further assume that the SGWBs satisfy
\begin{eqnarray}
\langle \tilde h_{A}(f,\hat{\bf n})\rangle=0 \ .
\end{eqnarray}
We also assume the directional dependence of the SGWB is frequency independent \cite{Allen:1996gp}. This implies the GW power spectral density is factorized into two parts, one of which depends on the direction while the other depends on the frequency.

Because of the transformations \erefs{henkan1}-\eqref{henkan2}, the parameters $I$ and $V$ have spin 0 and the parameters $Q\pm iU$ 
 have spin $\pm 4$~\cite{Seljak:1996gy}.
To analyze the SGWB on the sky, it is convenient to expand the Stokes parameters by spherical harmonics $Y_{lm}$.
However, since $Q\pm iU$ parameters have spin $\pm 4$, they have to be expanded by the spin-weighted harmonics $_{\pm4}Y_{lm}$ \cite{goldberg}.
Thus, we obtain
\begin{eqnarray}
I(f,\hat{\bf n})&=&I(f)\sum_{l=0}^{\infty}\sum_{m=-l}^{l}c^{I}_{lm}Y_{lm}(\hat {\bf n})\ ,\\
V(f,\hat{\bf n})&=&V(f)\sum_{l=0}^{\infty}\sum_{m=-l}^{l}c^{V}_{lm}Y_{lm}(\hat {\bf n})\ ,\\
(Q+iU)(f,\hat{\bf n})&=&P^{+}(f)\sum_{l=4}^{\infty}\sum_{m=-l}^{l}c^{P+}_{lm}{_{4}Y_{lm}}(\hat {\bf n})\ ,\\
(Q-iU)(f,\hat{\bf n})&=&P^{-}(f)\sum_{l=4}^{\infty}\sum_{m=-l}^{l}c^{P-}_{lm}{_{-4}Y_{lm}}(\hat {\bf n})\ .
\end{eqnarray}
In this paper, we study specific anisotropic patterns with $l=0,1,2,3$ for simplicity. 
Therefore, we can neglect $Q$ and $U$ from now on. Thus, the GW power spectral density becomes
\begin{eqnarray}
S_{h}^{AA'}(f,\hat{\bf n})=
\left(
    \begin{array}{cc}
       I_{3}(f,\hat{\bf n}) & -iV_{3}(f,\hat{\bf n})  \\
        iV_{3}(f,\hat{\bf n})  &  I_{3}(f,\hat{\bf n}) 
    \end{array}
\right) \ ,\label{sh3}
\end{eqnarray}
where
\begin{eqnarray}
I_{3}(f,\hat{\bf n})=I(f)\sum_{l=0}^{3}\sum_{m=-l}^{l}c^{I}_{lm}Y_{lm}(\hat {\bf n}) \ ,\quad
V_{3}(f,\hat{\bf n})=V(f)\sum_{l=0}^{3}\sum_{m=-l}^{l}c^{V}_{lm}Y_{lm}(\hat {\bf n}) \ .
\end{eqnarray}
So, we focus on the parameters $I$ and $V$ .
In what follows, we will use the following shorthand notation
\begin{eqnarray}
\sum_{lm}^{l=3}\equiv\sum_{l=0}^{3}\sum_{m=-l}^{l}.
\end{eqnarray}

Next, we consider the dimensionless density parameter~\cite{Maggiore-2008}
\begin{eqnarray}
\Omega_{\rm gw}(f)=\frac{1}{\rho_c}\frac{d\rho_{\rm gw}}{d\log f}\ ,\label{omega}
\end{eqnarray}
where $\rho_{c}=3H^2_0/8\pi$ is the critical density, $H_{0}$ is the present value of the Hubble parameter, 
\begin{eqnarray}
\rho_{\rm gw}=\frac{1}{32\pi}\langle\dot h_{ij}(t,{\bf x})\dot h^{ij}(t,{\bf x})\rangle \ ,\label{hdhd}
\end{eqnarray}
is the energy density of gravitational waves, 
and $d\rho_{gw}$ is the energy density in the frequency range $f$ to $f+df$.
The bracket $\langle ...\rangle$ represents the  ensemble average.
However, actually, we take a spatial average over the wave lengths $\lambda/(2\pi)$ of GWs 
or a temporal average over the periods $1/f$ of GWs.
Here, we assumed the ergodicity, namely, the ensemble average can be replaced by the temporal average. 
Using \erefs{hij}, \eqref{ee}, \eqref{pawa-}, as well as $\tilde h_{A}(-f,{\bf \hat n})=\tilde h_{A}^{\ast}(f,{\bf\hat n})$ and $I(f)=I(-f)$, we get
\begin{eqnarray}
\rho_{\rm gw}=\pi\int_{f=0}^{f=\infty} d(\log f)\,\,f^3I(f)\sum_{lm}\int_{S^2} d^2 {\bf\hat n}\,\,c^{I}_{lm}Y_{lm}({\bf\hat n}) \ .
\end{eqnarray}
Then we define 
\begin{eqnarray}
\rho_{\rm gw}\equiv\int_{f=0}^{f=\infty} d({\rm log} f)\frac{d\rho_{\rm gw}}{d{\rm log} f} \ .
\end{eqnarray}
Hence, the dimensionless quantity $\Omega_{\rm gw}(f)$ in \eref{omega} is given by
\begin{eqnarray}
\Omega_{\rm gw}(f)=\frac{8\pi^2}{3H_{0}^{2}} f^3I(f)\sum_{lm}\int_{S^2} d^2 {\bf\hat n}\,\,c^{I}_{lm}Y_{lm}({\bf\hat n}) \ ,
\end{eqnarray}
where the spherical harmonics are orthogonal and normalized as
\begin{eqnarray}
\int_{S^2} d^2 {\bf\hat n}\,\,Y_{lm}({\bf\hat n})Y_{l'm'}({\bf\hat n})=\delta_{ll'}\delta_{mm'} \ .
\end{eqnarray}
Using $Y_{00}({\bf\hat n})=1/ \sqrt{4\pi}$, we obtain
\begin{eqnarray}
\Omega_{\rm gw}(f)=\frac{16\pi^{5/2}c^{I}_{00}}{3H_{0}^{2}} f^3I(f) \ .\label{omegab}
\end{eqnarray}
Without loss of generality, we normalize the monopole moment as
\begin{eqnarray}
c^{I}_{00}=\sqrt{4\pi} \ .
\end{eqnarray}
So, \eref{omega} becomes
\begin{eqnarray}
\Omega_{\rm gw}(f)=\frac{32\pi^{3}}{3H_{0}^{2}} f^3I(f) \ .
\end{eqnarray}

\subsection{Frequency shift}\label{subsec:The signal}
The time of arrival of radio pulses from the pulsar is affected by GWs. 
Consider a pulsar with frequency $\nu_{0}$ located in the direction $\hat p$. 
To detect  the SGWB, let us consider the redshift of the pulse from a pulsar \cite{Anholm:2008wy,Book:2010pf}
\begin{eqnarray}
z(t,{\bf\hat n})   \equiv  \frac{\nu_{0}-\nu(t)}{\nu_{0}}  
   = \frac{1}{2}\frac{\hat p^{i}\hat p^{j}}{1+\hat{\bf n}\cdot\hat p} \Delta h_{ij} (t,\bf\hat n)            \ ,           \label{zto}
\end{eqnarray}
where $\nu(t)$ is a frequency detected at the Earth and $\hat p$ is the direction to the pulsar. 
The unit vector $\hat{\bf n}$ represents the direction of propagation of gravitational plane waves. We also defined
the difference between the metric perturbations at the pulsar $(t_{p},{{\bf x}_{p}})$ 
and at the Earth $(t_{e},{{\bf x}_{e}})$ as
\begin{eqnarray}
\Delta h_{ij}(t,{\bf\hat n})=h_{ij}(t_{p},\hat{\bf n})-h_{ij}(t_{e},\hat{\bf n})   \ .\label{dhij}
\end{eqnarray}
 The gravitational plane waves at each point is defined as
\begin{eqnarray}
h_{ij}(t,\hat{\bf n})\equiv\sum_{A=+,\times}\int_{-\infty}^{\infty} df \,\,
\tilde h_{A}(f,{\bf \hat n})\,\,e_{ij}^{A}(\hat{\bf n})\,\,\mathrm{e}^{{-2\pi if}(t-{\bf\hat n}\cdot{\bf x})}.
\end{eqnarray}
For the SGWB, the redshift have to be integrated over the direction of propagation of the gravitational waves $\hat{\bf n}$:
\begin{eqnarray}
z(t)=\int_{S^2} d^2 {\bf\hat n}\,\,z(t,\bf\hat n)\ .\label{zt}
\end{eqnarray}
We choose a coordinate system  
\begin{eqnarray}
t_{e} = t \ ,\quad
{\bf x}_{e} = 0 \ ,\quad
t_{p} = t_{e}-L=t-L \ ,\quad
{\bf x}_{p} = L\hat p \ ,
\end{eqnarray}
and assume that the amplitudes of the metric perturbation at the pulsar and the Earth are the same. Then \eref{dhij} becomes
\begin{eqnarray}
\Delta h_{ij}(t,{\bf\hat n})=\int_{-\infty}^{\infty}df\,\,\mathrm{e}^{-i2\pi ft}(\mathrm{e}^{2\pi ifL(1+\hat{\bf n}\cdot\hat{p})}-1)\sum_{A}h_{A}(f,\hat{\bf n})e_{ij}^{A}(\hat{\bf n})\ ,
\end{eqnarray}
and therefore, \eref{zt} reads 
\begin{eqnarray}
z(t)=\int_{-\infty}^{\infty}df\int_{S^2} d^2 {\bf\hat n}\,\,\mathrm{e}^{-i2\pi ft}(\mathrm{e}^{2\pi ifL(1+\hat{\bf n}\cdot\hat{p})}-1)\sum_{A}h_{A}(f,\hat{\bf n})F^{A}(\hat{\bf n})\ ,\label{zt2}
\end{eqnarray}
where we have defined the pattern functions for pulsars
\begin{eqnarray}
F^{A}(\hat{\bf n})\equiv\frac{1}{2}\frac{\hat p^{i}\hat p^{j}}{1+\hat{\bf n}\cdot\hat p}e_{ij}^{A}(\hat{\bf n}) \ .
\end{eqnarray}
Note that our convention for the Fourier transformation is
\begin{eqnarray}
g(t)=\int_{-\infty}^{\infty}df\,\,\tilde g(f)\mathrm{e}^{-i2\pi ft} \ .
\end{eqnarray} 
Therefore, the Fourier transformation of \eref{zt2} can be written as
\begin{eqnarray}
\tilde z(f)&=&\int_{S^2} d^2 {\bf\hat n}\,\,(\mathrm{e}^{2\pi ifL(1+\hat{\bf n}\cdot\hat{p})}-1)
\sum_{A}h_{A}(f,\hat{\bf n})F^{A}(\hat{\bf n}) \ .
\end{eqnarray}

In the actual signals from a pulsar, there exist noises. Hence, we need to use the correlation analysis. 
We consider the signals from two pulsars  
\begin{eqnarray}
s_{a}(t)=z_{a}(t)+n_{a}(t)\ ,
\end{eqnarray}
where $a=1,2$ labels the pulsar. Here, $s_{a}(t)$ denotes the signal from the pulsar and $n_{a}(t)$ denotes 
the noise intrinsic to the measurement. 
We assume the noises are stationary, Gaussian and are not correlated between the two pulsars.
To correlate the signals of two measurements, we define
\begin{eqnarray}
Y=\int_{-\frac{T}{2}}^{\frac{T}{2}}dt \int_{-\frac{T}{2}}^{\frac{T}{2}}dt'\,\,s_{1}(t)s_{2}(t')K(t-t') \ ,\label{y}
\end{eqnarray}
where $T$ is the total observation time and $K(t-t')$ is a real filter function which should be optimal to maximize signal-to-noise ratio.
In the case of interferometer, the optimal filter function falls to zero for large $|t-t'|$ compered to the travel time of the light  between the detecters.  Since the signals of two detectors are expected to correlate due to the same effect of the gravitational waves, 
the optimal filter function should  behave this way. Then, typically one of the detectors is very close to the other 
compared to the total observation time $T$.          
Therefore, the total observation time $T$ can be extended to $\pm \infty$ \cite{Allen:1997ad}. 
In contrast, in the case of PTA, it is invalid that $T$ is very large compered to the travel time of the light  between the pulsars.
Nevertheless, we can assume that one of the two $T$ can be expanded to $\pm \infty$, because in situations $L=L_{1}=L_{2}$ and $fL\gg1$ it is known that we can ignore the effect of the distance $L$ of pulsars.
In this case, it is clear that any locations of the pulsars are optimal and optimal filter function should behave like as the
 interferometer case \cite{Anholm:2008wy}. 

Using these assumptions $L=L_{1}=L_{2}$ and $fL\gg1$,  we can rewrite \eref{y} as
\begin{eqnarray}
Y=\int_{-\infty}^{\infty}\,\,df\int_{-\infty}^{\infty}\,\,df'\,\,\delta_{T}(f-f')\tilde s^{\ast}_{1}(f)\tilde s_{2}(f')\tilde K(f')\ ,
\end{eqnarray}
where 
\begin{eqnarray}
\delta_{T}(f)  \equiv     \int_{-\frac{T}{2}}^{\frac{T}{2}}dt\,\,\mathrm{e}^{2\pi ift}
=\frac{\sin(\pi fT)}{\pi f} \ .
\end{eqnarray}
Note that $\tilde K(f)$ satisfies $\tilde K(f)=\tilde K^{\ast}(-f)$, because $K(t)$ is real.
 Moreover, to deal with the unphysical region $f\le0$ we require $\tilde K(f)=\tilde K(-f)$.
 Thus, $\tilde K(f)$ becomes real.
Taking the ensemble average,  using $\delta(0)=\int_{-T/2}^{T/2}dt=T$, $L=L_{1}=L_{2}$, and assuming the noises in the two measurements
 are not correlated, we get 
\begin{eqnarray}
S&\equiv&\langle Y\rangle   
 = \int_{-\infty}^{\infty}\,\,df\int_{-\infty}^{\infty}\,\,df'\,\,\delta_{T}(f-f')\langle\tilde z^{\ast}_{1}(f)\tilde z_{2}(f')\rangle\tilde K(f')\label{s1}     \nonumber \\
&=&T\int_{-\infty}^{\infty} df\,\,\tilde K(f)\left[I(f)\Gamma^{I}(f)+V(f)\Gamma^{V}(f)\right]\ ,\label{s2}
\end{eqnarray}
where we have defined 
\begin{eqnarray}
\Gamma^{I}(f)&\equiv&\int_{S^2}d\hat{\bf n}\,\,\kappa(f,\hat{\bf n})\sum_{lm}^{l=3}c_{lm}^{I} Y_{lm}(\hat {\bf n})\left\{F_{1}^{\ast +}(\hat{\bf n})F_{2}^{+}(\hat{\bf n})+F_{1}^{\ast \times}(\hat{\bf n})F_{2}^{\times}(\hat{\bf n})\right\} \ ,\\
\Gamma^{V}(f)&\equiv&-i\int_{S^2}d\hat{\bf n}\,\,\kappa(f,\hat{\bf n})\sum_{lm}^{l=3}c_{lm}^{V} Y_{lm}(\hat {\bf n})\left\{F_{1}^{\ast +}(\hat{\bf n})F_{2}^{\times}(\hat{\bf n})-F_{1}^{\ast \times}(\hat{\bf n})F_{2}^{+}(\hat{\bf n})\right\} \ ,\\
\kappa(f,\hat{\bf n})&\equiv&(\mathrm{e}^{{-2\pi ifL}(1+\hat{\bf n}\cdot\hat{p}_1)}-1)(\mathrm{e}^{{2\pi ifL}(1+\hat{\bf n}\cdot\hat{p}_2)}-1) \ .\label{kappa}
\end{eqnarray}
The functions $\Gamma^{I}(f)$ and $\Gamma^{V}(f)$ are called the generalized ORFs, which describe the angular sensitivity of the pulsars for the SGWB. Note that, as we already mentioned, we consider the cases of $l=0,1,2,3$ for simplicity. Then we have assumed $L=L_{1}=L_{2}$ and $fL\gg1$, this assumption implies that \eref{kappa}  approximately becomes
\begin{eqnarray}
\kappa(f,\hat{\bf n})\sim1 \ ,\label{kappa2}
\end{eqnarray}
due to the rapid oscillation of the phase factor. 
Therefore,  the distance $L$ of the pulsars does not appear in the generalized ORFs, 
and hence the generalized ORFs do not depend on the frequency.

As you can see from \eref{s2}, the correlation of the two measurements involve both the total intensity and the 
circular polarization. 
However, the degeneracy can be disentangled by using separation method, which will be discussed in the section \ref{sec:separation method}.

\section{Generalized overlap reduction function for circular polarization}\label{sec:the generalized overlap reduction function for circular polarization}
In this section, we consider the generalized ORFs for circular polarizations:
\begin{eqnarray}
\Gamma^{V}=\sum_{lm}^{l=3} c_{lm}^{V}\Gamma^{V}_{lm} \ ,
\end{eqnarray}
where we defined 
\begin{eqnarray}
\Gamma^{V}_{lm}\equiv-i\int_{S^2}d\hat{\bf n}\,\,Y_{lm}(\hat {\bf n})\left\{F_{1}^{\ast +}(\hat{\bf n})F_{2}^{\times}(\hat{\bf n})-F_{1}^{\ast \times}(\hat{\bf n})F_{2}^{+}(\hat{\bf n})\right\} \ .
\end{eqnarray}
In the above, we have used \eref{kappa2} and the fact that the generalized ORFs do not depend on frequency.
For computation of the generalized ORFs for circular polarizations, it is convenient 
to use the computational frame \cite{Mingarelli:2013dsa} defined by 
\begin{eqnarray}
\hat p_{1}&=&(0,0,1)\ ,\label{p1}\\
\hat p_{2}&=&(\sin\xi,0,\cos\xi)\label{p2}\ ,
\end{eqnarray}
where $\xi$ is the angular separation between the two pulsars. 
Using \erefs{kika1}-\eqref{kika2}, \eqref{p1}, and \eqref{p2}, one can easily show that
\begin{eqnarray}
F_{1}^{+}(\hat{\bf n})&=&\frac{1}{2}\frac{\hat p_{1}^{i}\hat p_{1}^{j}}{1+\hat{\bf n}\cdot\hat p_{1}}e_{ij}^{+}(\hat{\bf n})=-\frac{1}{2}(1-\cos\theta)\ ,\\
F_{1}^{\times}(\hat{\bf n})&=&\frac{1}{2}\frac{\hat p_{1}^{i}\hat p_{1}^{j}}{1+\hat{\bf n}\cdot\hat p_{1}}e_{ij}^{\times}(\hat{\bf n})=0\ ,\\
F_{2}^{+}(\hat{\bf n})&=&\frac{1}{2}\frac{\hat p_{2}^{i}\hat p_{2}^{j}}{1+\hat{\bf n}\cdot\hat p_{2}}e_{ij}^{+}(\hat{\bf n})=\frac{1}{2}\frac{\sin^2\xi\sin^2\phi-\sin^2\xi\cos^2\theta\cos^2\phi+2\sin\xi\cos\xi\sin\theta\cos\theta\cos\phi-\cos^2\xi\sin^2\theta}{1+\sin\xi\sin\theta\cos\phi+\cos\xi\cos\theta}\ ,\\
F_{2}^{\times}(\hat{\bf n})&=&\frac{1}{2}\frac{\hat p_{2}^{i}\hat p_{2}^{j}}{1+\hat{\bf n}\cdot\hat p_{2}}e_{ij}^{\times}(\hat{\bf n})=\frac{\sin^2\xi\cos\theta\sin\phi\cos\phi-\sin\xi\cos\xi\sin\theta\sin\phi}{1+\sin\xi\sin\theta\cos\phi+\cos\xi\cos\theta} \ .
\end{eqnarray}
We therefore get
\begin{eqnarray}
\Gamma^{V}_{lm}(f)=\frac{i}{2}\int_{S^2}d\hat{\bf n}\,\,Y_{lm}(\hat{\bf n})\frac{(1-\cos\theta)(\sin^2\xi\cos\theta\sin\phi\cos\phi-\sin\xi\cos\xi\sin\theta\sin\phi)}{1+\sin\xi\sin\theta\cos\phi+\cos\xi\cos\theta} \ .\label{gvlm1}
\end{eqnarray}
The explicit form of the spherical harmonics reads
\begin{eqnarray}
Y_{lm}(\theta,\phi)=N_{l}^{m}P_{l}^{m}(\cos\theta)\mathrm{e}^{im\phi}\label{ylm} \ ,
\end{eqnarray}
where
\begin{eqnarray}
N_{l}^{m}&=&\sqrt{\frac{2l+1}{4\pi}\frac{(l-m)!}{(l+m)!}}\ ,   \label{nlm}
\end{eqnarray}
is the normalization factor. The associated Legendre functions are given by
\begin{eqnarray}
P_{l}^{m}(x) = (-1)^{m}(1-x^2)^{m/2} \frac{d^{m}}{dx^{m}}P_{l}(x) \ ,
\end{eqnarray}
and
\begin{eqnarray}
P_{l}^{-m}(x) = (-1)^{m}\frac{(l-m)!}{(l+m)!}P_{l}^{m}(x) \ ,
\end{eqnarray}
with the Legendre functions
\begin{eqnarray}
P_{l}&=&\frac{1}{2^{l}l!}\frac{d^{l}}{dx^{l}}[(x^2-1)^{l}]\ .\label{pl}
\end{eqnarray}
 Using the spherical harmonics, \eref{gvlm1} becomes
\begin{eqnarray}
\Gamma^{V}_{lm}
&=&-\frac{N_{l}^{m}}{2}\int_{0}^{\pi}d\theta\,\,\sin\theta (1-\cos\theta)P_{l}^{m}(\cos\theta)\int_{0}^{2\pi}d\phi\,\,\frac{(\sin^2\xi\cos\theta\sin\phi\cos\phi-\sin\xi\cos\xi\sin\theta\sin\phi)}{1+\sin\xi\sin\theta\cos\phi+\cos\xi\cos\theta}\sin(m\phi)\,\,\,,\notag\\\label{gvlm2}
\end{eqnarray}
where we have used the fact that the function of $\phi$ is odd parity in the case of $\cos m\phi$ 
and is even parity in the case of $i\sin m\phi$.
Note that the generalized ORFs for circular polarizations are real functions. In the case of $m=0$ and/or $\xi=0, \pi$, 
 the integrand in \eref{gvlm2} vanishes. Therefore, we cannot detect circular polarizations for these cases. 
This fact for $\xi=0$ implies that we do not need to consider auto-correlation for a single pulsar.
This is the reason why we neglected auto-correlation term in \eref{kappa2}.

Integrating \eref{gvlm2}, we get the following form for $l=0$:
\begin{eqnarray}
\Gamma^{V}_{00}&=&0 \ .
\end{eqnarray}
For $l=1$, we have obtained 
\begin{eqnarray}
\Gamma^{V}_{10}&=&0 \ ,\\
\Gamma^{V}_{11}&=&-\frac{\sqrt{6\pi}}{3}
\sin\xi\left[1+3\left(\frac{1-\cos\xi}{1+\cos\xi}\right)\log\left(\sin\frac{\xi}{2}\right)\right] \ ,\\
\Gamma^{V}_{1-1}&=&\Gamma^{V}_{11} \ ,
\end{eqnarray}
 Recall that $-l\le m\le l$. 
 The derivation of this formula for $l=1$ can be found
 in Appendix \ref{sec:angular integral of the generalized overlap reduction function for dipole circular polarization}.

\begin{figure}[t]
\begin{minipage}{0.4\hsize}
\begin{center}
\includegraphics[width=9.0cm,bb=0 0 700 400]{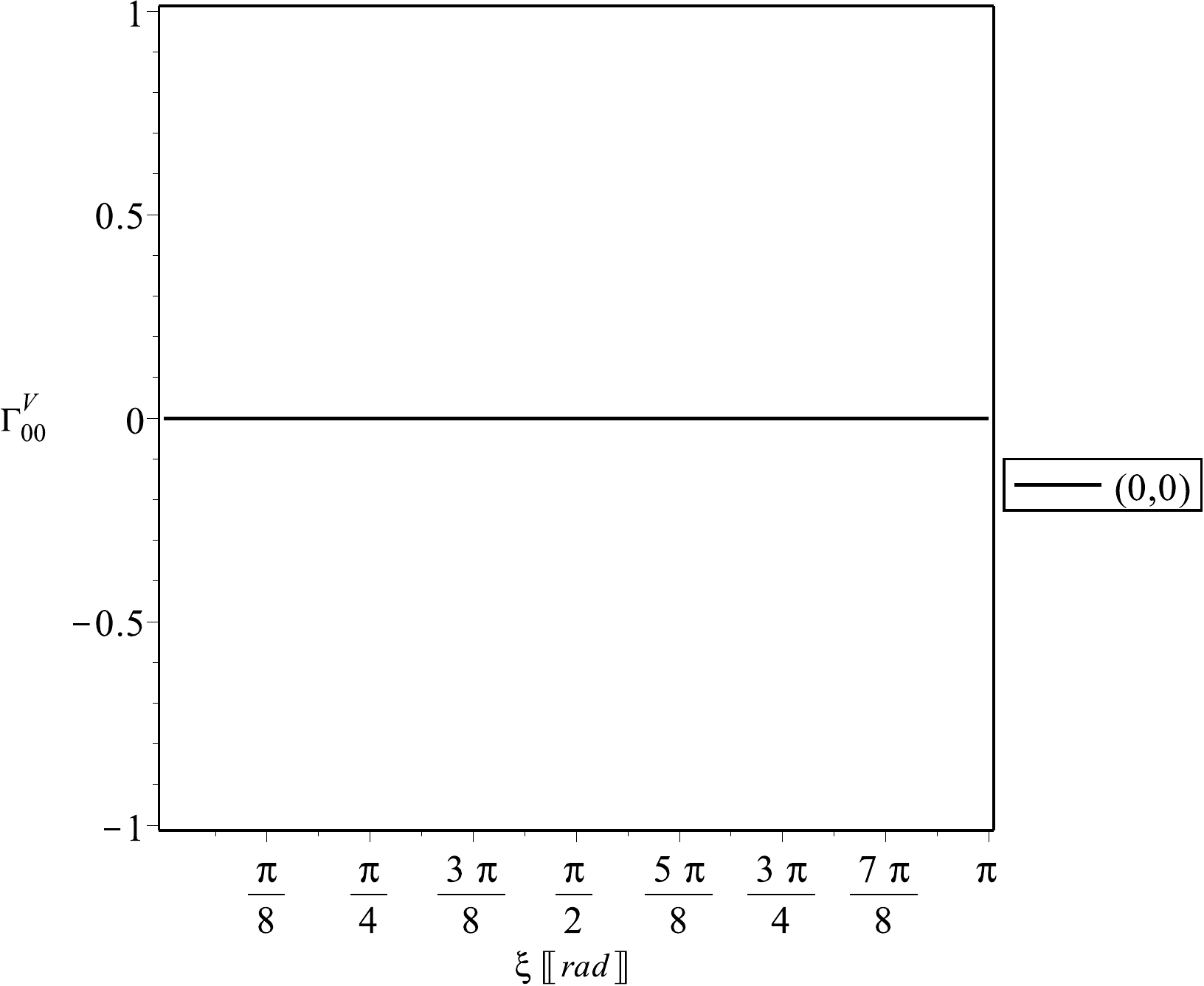}
\hspace{1.6cm} (a) $l=0$
\end{center}
\end{minipage}
\begin{minipage}{0.4\hsize}
\begin{center}
\includegraphics[width=9.0cm,bb=0 0 700 400]{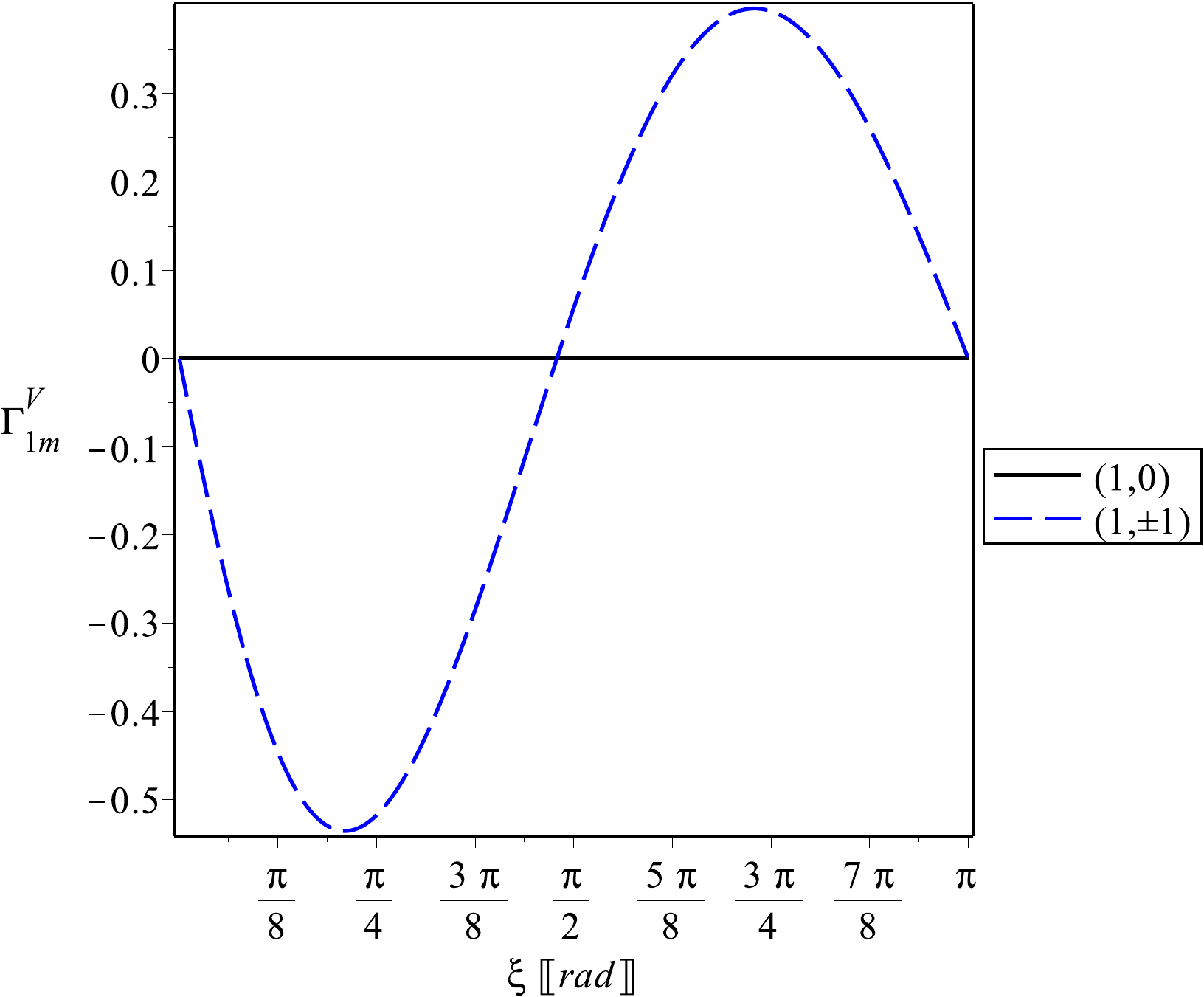}
\hspace{1.6cm} (b) $l=1$
\end{center}
\end{minipage}
\begin{minipage}{0.4\hsize}
\begin{center}
\includegraphics[width=9.0cm,bb=0 0 700 400]{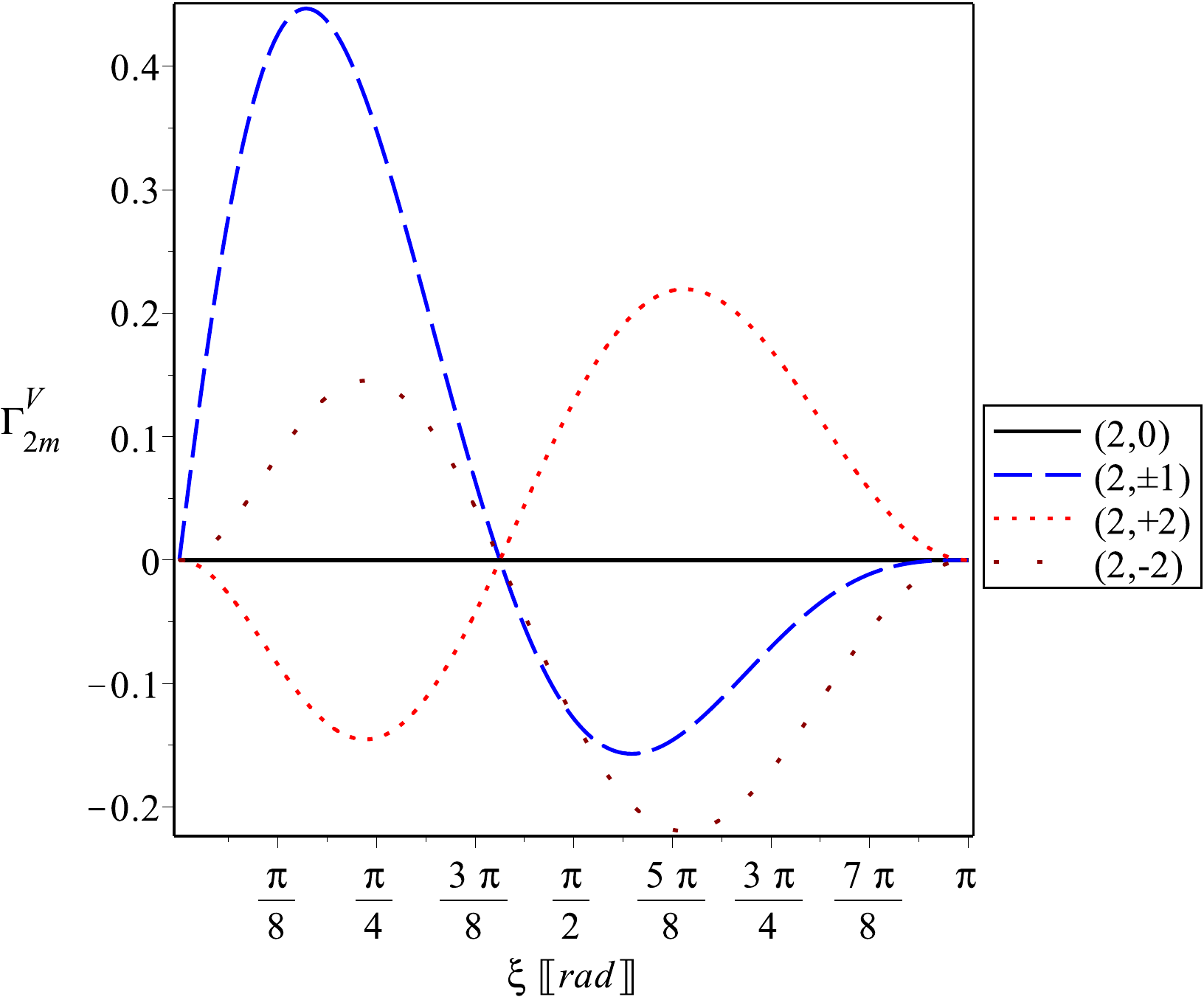}
\hspace{1.6cm} (c) $l=2$
\end{center}
\end{minipage}
\begin{minipage}{0.4\hsize}
\begin{center}
\includegraphics[width=9.0cm,bb=0 0 700 400]{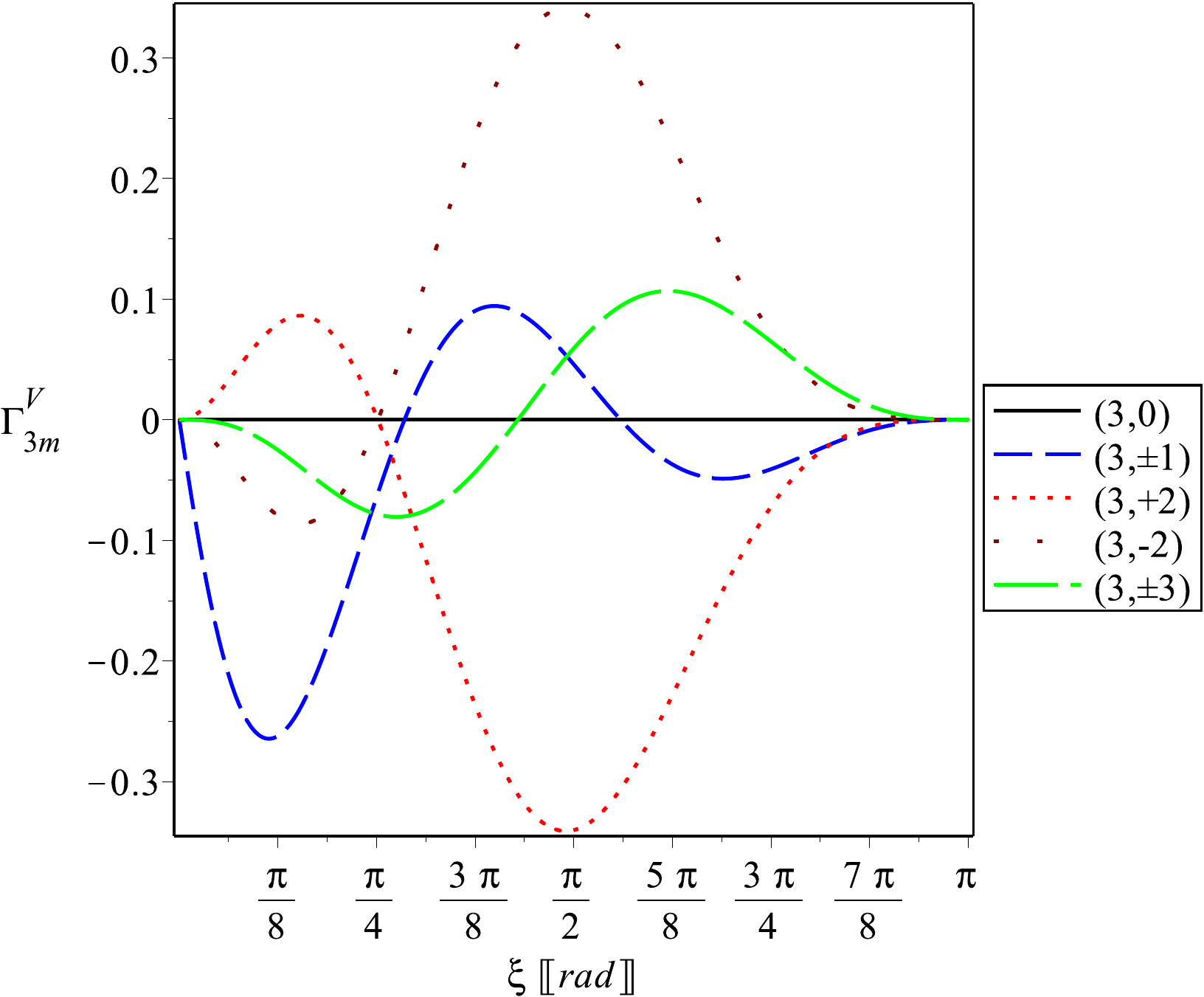}
\hspace{1.6cm} (d) $l=3$
\end{center}
\end{minipage}
\caption{Plots of the generalized ORFs $\Gamma_{lm}^{V}$ as a function of the angular separation between the two pulsars $\xi$.
In Fig. \ref{GV}(a),  we find the ORF for the monopole (l=0) is trivial. In Fig. \ref{GV}(b), the ORFs for the dipole (l=1) are shown.
 In Fig. \ref{GV}(c), the ORFs for the quadrupole (l=2) are depicted. In  Fig. \ref{GV}(d), the ORFs for the octupole (l=3) are plotted.
The black solid curve, the blue dashed curve, the red dotted curve, the dark-red space-dotted curve, and the green long-dashed curve  
represent $m=0$, $m=\pm1$, $m=+2$, $m=-2$, $m=\pm3$, respectively.}
\label{GV}
\end{figure}

For $l=2$, we derived the following:
\begin{eqnarray}
\Gamma^{V}_{20}&=&0\ ,\\
\Gamma^{V}_{21}&=&\frac{\sqrt{30\pi}}{6}\frac{\sin\xi}{1+\cos\xi}\left[2+(1-\cos\xi)\left\{\cos\xi+6\log\left(\sin\frac{\xi}{2}\right)\right\}\right]\ ,\\
\Gamma^{V}_{2-1}&=&\Gamma^{V}_{21}\ ,\\
\Gamma^{V}_{22}&=&-\frac{\sqrt{30\pi}}{6}(1-\cos\xi)\left[2-\cos\xi+6\left(\frac{1-\cos\xi}{1+\cos\xi}\right)\log\left(\sin\frac{\xi}{2}\right)\right]\ ,\\
\Gamma^{V}_{2-2}&=&-\Gamma^{V}_{22}\ ,
\end{eqnarray}
For $l=3$, the results are
\begin{eqnarray}
\Gamma^{V}_{30}&=&0\ ,\\
\Gamma^{V}_{31}&=&-\frac{\sqrt{21\pi}}{48}\sin\xi\left[33-20\cos\xi-5\cos^2\xi+96\left(\frac{1-\cos\xi}{1+\cos\xi}\right)\log\left(\sin\frac{\xi}{2}\right)\right]\ ,\\
\Gamma^{V}_{3-1}&=&\Gamma^{V}_{31}\ ,\\
\Gamma^{V}_{32}&=&\frac{\sqrt{210\pi}}{24}(1-\cos\xi)\left[8-5\cos\xi-\cos^2\xi+24\left(\frac{1-\cos\xi}{1+\cos\xi}\right)\log\left(\sin\frac{\xi}{2}\right)\right]\ ,\\
\Gamma^{V}_{3-2}&=&-\Gamma^{V}_{3-2}\ ,\\
\Gamma^{V}_{33}&=&-\frac{\sqrt{35\pi}}{16}\sin\xi\left(\frac{1-\cos\xi}{1+\cos\xi}\right)\left[11-6\cos\xi-\cos^2\xi+32\left(\frac{1-\cos\xi}{1+\cos\xi}\right)\log\left(\sin\frac{\xi}{2}\right)\right]\ ,\\
\Gamma^{V}_{3-3}&=&\Gamma^{V}_{33}\ .
\end{eqnarray}
In Fig. \ref{GV}, we plotted  these generalized ORFs as a function of the angular separation between the two pulsars $\xi$.

It is apparent that considering the $V$ mode does not make sense when we only consider the isotropic ($l=0$) ORF.
On the other hand, when we consider anisotropic ($l\neq0$) ORFs, it is worth taking into account polarizations.
The polarizations of the SGWB would give us rich information both  of
 super massive black hole binaries and of inflation in the early universe.
 
Using the same procedure described in the above to derive the generalized ORFs for circular polarizations, 
we can also derive the generalized ORFs for the intensity
\begin{eqnarray}
\Gamma^{I}=\sum_{lm}^{l=3} c_{lm}^{I}\Gamma^{I}_{lm} \ ,
\end{eqnarray}
where
\begin{eqnarray}
\Gamma^{I}_{lm}\equiv\int_{S^2}d\hat{\bf n}\,\,Y_{lm}(\hat {\bf n})\left\{F_{1}^{\ast +}(\hat{\bf n})F_{2}^{+}(\hat{\bf n})+F_{1}^{\ast \times}(\hat{\bf n})F_{2}^{\times}(\hat{\bf n})\right\} \ .
\end{eqnarray}
The angular integral in this case was performed in \cite{Mingarelli:2013dsa}. 
The results are summarized in Appendix \ref{sec:the generalized overlap reduction function for intensity}.

\section{Extraction of circularly polarized components}\label{sec:separation method}
In this section, we separate the $I$ mode and $V$ mode of the SGWB with correlation analysis \cite{Seto:2008sr}.
To this aim, we use four pulsars (actually we need at least three pulsars), and define correlations of $\tilde z(f)$
\begin{eqnarray}
&&\langle\tilde z_{1}^{\ast}(f)\tilde z_{2}(f')\rangle\equiv C_{12}(f)\delta(f-f') \ ,\\
&&\langle\tilde z_{3}^{\ast}(f)\tilde z_{4}(f')\rangle\equiv C_{34}(f)\delta(f-f') \ ,
\end{eqnarray}
where $1,2,3,4$ label the pulsars. Comparing \erefs{s1} with \eqref{s2},  we obtain
\begin{eqnarray}
&&C_{12}(f)=\sum_{lm}^{l=3}\left[c_{lm}^{I}I(f)\Gamma_{lm,12}^{I}+c_{lm}^{V}V(f)\Gamma_{lm,12}^{V}\right] \ ,\label{1c12}\\
&&C_{34}(f)=\sum_{lm}^{l=3}\left[c_{lm}^{I}I(f)\Gamma_{lm,34}^{I}+c_{lm}^{V}V(f)\Gamma_{lm,34}^{V}\right] \ .\label{1c34}
\end{eqnarray}
If the $I$ mode and $V$ mode of the SGWB are dominated by a certain $ l, m$ and $ l', m'$, 
 \erefs{1c12} and \eqref{1c34} become
\begin{eqnarray}
&&C_{12}(f)=\left[c_{ l m}^{I}I(f)\Gamma_{ l m,12}^{I}+c_{ l' m'}^{V}V(f)\Gamma_{ l' m',12}^{V}\right] \ ,\label{2c12} \\
&&C_{34}(f)=\left[c_{ l m}^{I}I(f)\Gamma_{ l m,34}^{I}+c_{ l' m'}^{V}V(f)\Gamma_{ l' m',34}^{V}\right] \ .\label{2c34}
\end{eqnarray}
 To separate the intensity and the circular polarization, we take the following linear combinations
\begin{eqnarray}
D_{I}&\equiv&a_{I}C_{34}(f)+b_{I}C_{12}(f)\label{di} \nonumber\\
&=&c_{{l}{m}}^{I}I(f)\left(\Gamma_{{l}{m},34}^{I}\Gamma^{V}_{{l}'{m}',12}-\Gamma_{{l}{m},12}^{I}\Gamma^{V}_{{l}'{m}',34}\right)\ ,\\
D_{V}&\equiv&a_{V}C_{34}(f)+b_{V}C_{12}(f)\label{dv}  \nonumber\\
&=&c_{{l}'{m}'}^{V}V(f)\left(\Gamma_{{l}'{m}',34}^{V}\Gamma^{I}_{{l}{m},12}-\Gamma_{{l}'{m}',12}^{V}\Gamma^{I}_{{l}{m},34}\right)\ ,
\end{eqnarray}
where we defined coefficients
\begin{eqnarray}
a_{I} \equiv \Gamma^{V}_{{l}'{m}',12} \label{ai} \ , \quad
b_{I} \equiv  -\Gamma^{V}_{{l}'{m}',34} \label{bi} \ , \quad
a_{V} \equiv \Gamma^{I}_{{l}{m},12} \label{av} \ ,\quad
b_{V} \equiv   -\Gamma^{I}_{{l}{m},34} \label{bv}  \ .
\end{eqnarray}
As you can see, $D_{I}$ contains only $I(f)$, and $D_{V}$ contains only $V(f)$.

For the signal $S_{p}$, the formulas corresponding to \erefs{di} and \eqref{dv} are given by
\begin{eqnarray}
S_{P} = \langle a_{P}Y_{34}+b_{P}Y_{12}\rangle   
 = T\int_{-\infty}^{\infty} df\,\,\tilde K(f)[a_{P}C_{34}(f)+b_{P}C_{12}(f)] \ ,\label{sp}
\end{eqnarray}
where $P$ denotes $I$ and $V$. We assume
\begin{eqnarray}
z(t)\ll n(t) \ ,
\end{eqnarray}
and that the noise in the four pulsars are not correlated.
We also assume that the ensemble average of Fourier amplitudes of the noises $\tilde n(f)$ is of the form
\begin{eqnarray}
\langle \tilde n^{\ast}(f)\tilde n(f')\rangle=\delta(f-f')S_{n}(f) \ ,
\end{eqnarray}
where $S_{n}(f)$ is the noise power spectral density.
 The reality of $n(t)$ gives rise to $\tilde n(-f)=\tilde n^{\ast}(f)$ and therefore 
 we obtain $S_{n}(-f)=S_{n}(f)$.  Without loss of generality, we can assume  
\begin{eqnarray}
\langle \tilde n(f)\rangle=0 \ .
\end{eqnarray}
Then we obtain corresponding noises $N_{P}^{2}$:
\begin{eqnarray}
N^{2}_{P} \equiv \langle (a_{P}Y_{34}+b_{P}Y_{12})^2\rangle-\langle a_{P}Y_{34}+b_{P}Y_{12}\rangle^2  
 = T\int_{-\infty}^{\infty} df\,\,|\tilde K(f)|^2[a^2_{P}S^{2}_{n,34}(f)+b^2_{P}S^{2}_{n,12}(f)]\ ,\label{np}
\end{eqnarray}
where 
\begin{eqnarray}
S_{n,12}(f) \equiv [S_{n,1}(f)S_{n,2}(f)]^{1/2} \label{sn12} \ , \quad
S_{n,34}(f) \equiv [S_{n,3}(f)S_{n,4}(f)]^{1/2} \label{sn34} \ .
\end{eqnarray}
Using the inner product
\begin{eqnarray}
(A,B)_{P}&\equiv&\int_{\infty}^{\infty}df\,\,A^{\ast}(f)B(f)[a^2_{P}S^{2}_{n,34}(f)+b^2_{P}S^{2}_{n,12}(f)] \ ,
\end{eqnarray}
we can rewrite \erefs{sp}, \eqref{np}  as
\begin{eqnarray}
S_{P}&=&T\left(\tilde K(f),\frac{a_{P}C_{34}(f)+b_{P}C_{12}(f)}{a^2_{P}S^{2}_{n,34}(f)+b^2_{P}S^{2}_{n,12}(f)}\right)_{P} \ ,\\
N_{P}&=& T^{1/2}(\tilde K(f),\tilde K(f))^{1/2}_{P} \ .
\end{eqnarray}
Therefore, the optimal filter function can be chosen as
\begin{eqnarray}
\tilde K_{P}(f)&=&
\frac{a_{P}C_{34}(f)+b_{P}C_{12}(f)}{a^2_{P}S^{2}_{n,34}(f)+b^2_{P}S^{2}_{n,12}(f)}\ .\label{kp}
\end{eqnarray}
Using \eref{kp}, we get optimal signal-to-noise ratio 
\begin{eqnarray}
{\rm SNR}_{P}
&\equiv& \frac{S_{P}}{N_{P}}
 = T^{1/2}\left(\frac{a_{P}C_{34}(f)+b_{P}C_{12}(f)}{a^2_{P}S^{2}_{n,34}(f)+b^2_{P}S^{2}_{n,12}(f)},\frac{a_{P}C_{34}(f)+b_{P}C_{12}(f)}{a^2_{P}S^{2}_{n,34}(f)+b^2_{P}S^{2}_{n,12}(f)}\right)^{1/2}_{P} \nonumber\\
&=&\left[T\int_{-\infty}^{\infty}df\,\,\frac{(a_{P}C_{34}(f)+b_{P}C_{12}(f))^2}{a^2_{P}S^{2}_{n,34}(f)+b^2_{P}S^{2}_{n,12}(f)}\right]^{1/2}\ .\label{snr}
\end{eqnarray}
Plugging \erefs{2c12}, \eqref{2c34}, and \eqref{ai} into \eref{snr}, we obtain
\begin{eqnarray}
{\rm SNR}_{I}&=&\left[T\int_{-\infty}^{\infty}df\,\,\frac{\left(c^{I}_{{l}{m}}\right)^{2}I^{2}(f)\left(\Gamma_{{l}{m},34}^{I}\Gamma^{V}_{{l}'{m}',12}-\Gamma_{{l}{m},12}^{I}\Gamma^{V}_{{l}'{m}',34}\right)^2}{\left(\Gamma^{V}_{{l}'{m}',12}\right)^2S^{2}_{n,34}(f)+\left(\Gamma^{V}_{{l}'{m}',34}\right)^2S^{2}_{n,12}(f)}\right]^{1/2}\ ,\\
{\rm SNR}_{V}&=&\left[T\int_{-\infty}^{\infty}df\,\,\frac{\left(c^{V}_{{l}'{m}'}\right)^{2}V^{2}(f)\left(\Gamma_{{l}'{m}',34}^{V}\Gamma^{I}_{{l}{m},12}-\Gamma_{{l}'{m}',12}^{V}\Gamma^{I}_{{l}{m},34}\right)^2}{\left(\Gamma^{I}_{{l}{m},12}\right)^2S^{2}_{n,34}(f)+\left(\Gamma^{I}_{{l}{m},34}\right)^2S^{2}_{n,12}(f)}\right]^{1/2}\ .
\end{eqnarray} 
If we assume all of the noise power spectral densities are the same, \eref{sn12}  becomes
\begin{eqnarray}
S_{n,12}(f)=S_{n,34}(f) \ .
\end{eqnarray}
Thus,  the compiled ORFs can be defined as
\begin{eqnarray}
\Gamma_{12:34}^{I}&\equiv&\frac{\Gamma_{{l}{m},34}^{I}\Gamma^{V}_{{l}'{m}',12}-\Gamma_{{l}{m},12}^{I}\Gamma^{V}_{{l}'{m}',34}}{\left[\left(\Gamma^{V}_{{l}'{m}',12}\right)^2+\left(\Gamma^{V}_{{l}'{m}',34}\right)^2\right]^{1/2}}\ ,\\
\Gamma_{12:34}^{V}&\equiv&\frac{\Gamma_{{l}'{m}',34}^{V}\Gamma^{I}_{{l}{m},12}-\Gamma_{{l}'{m}',12}^{V}\Gamma^{I}_{{l}{m},34}}{\left[\left(\Gamma^{I}_{{l}{m},12}\right)^2+\left(\Gamma^{I}_{{l}{m},34}\right)^2\right]^{1/2}}\ .
\end{eqnarray}
This compiled ORFs $\Gamma_{{l}{m}{l}'{m}',12:34}^{I}$ and $\Gamma_{{l}{m}{l}'{m}',12:34}^{V}$ describe the angular sensitivity of the four pulsars for the pure $I$ and $V$ mode of the SGWB, respectively.
Note that, to do this separation, we must know a priori the coefficients $a_{P}$ and $b_{P}$.
If we do not assume \eref{kappa2}, the generalized ORFs depend on the frequency. 
In this case, it seems difficult to calculate these coefficients. 

We next consider the case that $I$ mode and/or $V$ mode dominant in two or more $l,m$.
In this case, if we have a priori knowledge of the values of $c^{P}_{lm}$ in each of $l,m$ for coefficients $a_{P}$ and $b_{P}$, we can separate $I$ mode and $V$ mode.
For example, assume that $I$ mode is dominated by $(l, m)=(0,0),(1,1)$, while $V$ mode is dominated by $( l', m')=(1,1)$,
 then \erefs{1c12} and \eqref{1c34} become
\begin{eqnarray}
&&C_{12}(f)=\left[c^{I}_{00}I(f)\left(\Gamma_{00,12}^{I}+\frac{c^{I}_{11}}{c^{I}_{00}}\Gamma_{11,12}^{I}\right)+c_{11}^{V}V(f)\Gamma_{11,12}^{V}\right]\ ,\label{3c12}\\
&&C_{34}(f)=\left[c^{I}_{00}I(f)\left(\Gamma_{00,34}^{I}+\frac{c^{I}_{11}}{c^{I}_{00}}\Gamma_{11,34}^{I}\right)+c_{11}^{V}V(f)\Gamma_{11,34}^{V}\right]\ .\label{3c34}
\end{eqnarray}
Thus, we can separate $I$ mode and $V$ mode by using linear combinations
\begin{eqnarray}
D_{I}&\equiv&a_{I}C_{34}(f)+b_{I}C_{12}(f) \nonumber\\
&=&c^{I}_{00}I(f)\left[\left(\Gamma_{00,34}^{I}+\frac{c^{I}_{11}}{c^{I}_{00}}\Gamma_{11,34}^{I}\right)\Gamma^{V}_{11,12}-\left(\Gamma_{00,12}^{I}+\frac{c^{I}_{11}}{c^{I}_{00}}\Gamma_{11,12}^{I}\right)\Gamma^{V}_{11,34}\right]\ ,\\
D_{V}&\equiv&a_{V}C_{34}(f)+b_{V}C_{12}(f) \nonumber\\
&=&c_{11}^{V}V(f)\left[\Gamma_{11,34}^{V}\left(\Gamma_{00,12}^{I}+\frac{c^{I}_{11}}{c^{I}_{00}}\Gamma_{11,12}^{I}\right)-\Gamma_{11,12}^{V}\left(\Gamma_{00,34}^{I}+\frac{c^{I}_{11}}{c^{I}_{00}}\Gamma_{11,34}^{I}\right)\right]\ ,
\end{eqnarray}
where
\begin{eqnarray}
a_{I} \equiv \Gamma^{V}_{11,12}\label{2ai}  \ ,\quad
b_{I} \equiv -\Gamma^{V}_{11,34}\label{2bi}\ ,\quad
a_{V} \equiv \Gamma_{00,12}^{I}+\frac{c^{I}_{11}}{c^{I}_{00}}\Gamma_{11,12}^{I}\label{2av}\ ,\quad
b_{V} \equiv -\Gamma_{00,34}^{I}-\frac{c^{I}_{11}}{c^{I}_{00}}\Gamma_{11,34}^{I}\ .\label{2bv}
\end{eqnarray}
As in the previous calculations, we can get the compiled ORFs
\begin{eqnarray}
\Gamma_{12:34}^{I}&\equiv&\frac{\left(\Gamma_{00,34}^{I}+\displaystyle\frac{c^{I}_{11}}{c^{I}_{00}}\Gamma_{11,34}^{I}\right)\Gamma^{V}_{11,12}-\left(\Gamma_{00,12}^{I}+\displaystyle\frac{c^{I}_{11}}{c^{I}_{00}}\Gamma_{11,12}^{I}\right)\Gamma^{V}_{11,34}}{\left[\left(\Gamma^{V}_{11,12}\right)^2+\left(\Gamma^{V}_{11,34}\right)^2\right]^{1/2}}\ ,\label{gi1234}\\
\Gamma_{12:34}^{V}&\equiv&\frac{\Gamma_{11,34}^{V}\left(\Gamma_{00,12}^{I}+\displaystyle\frac{c^{I}_{11}}{c^{I}_{00}}\Gamma_{11,12}^{I}\right)-\Gamma_{11,12}^{V}\left(\Gamma_{00,34}^{I}+\displaystyle\frac{c^{I}_{11}}{c^{I}_{00}}\Gamma_{11,34}^{I}\right)}{\left[\left(\Gamma_{00,12}^{I}+\displaystyle\frac{c^{I}_{11}}{c^{I}_{00}}\Gamma_{11,12}^{I}\right)^2+\left(\Gamma_{00,34}^{I}+\displaystyle\frac{c^{I}_{11}}{c^{I}_{00}}\Gamma_{11,34}^{I}\right)^2\right]^{1/2}}\ .\label{gv1234}
\end{eqnarray}

\begin{figure}[pt]
\begin{tabular}{cc}
\begin{minipage}{0.31\hsize}
\begin{center}
\includegraphics[width=9.0cm,bb=0 0 700 400]{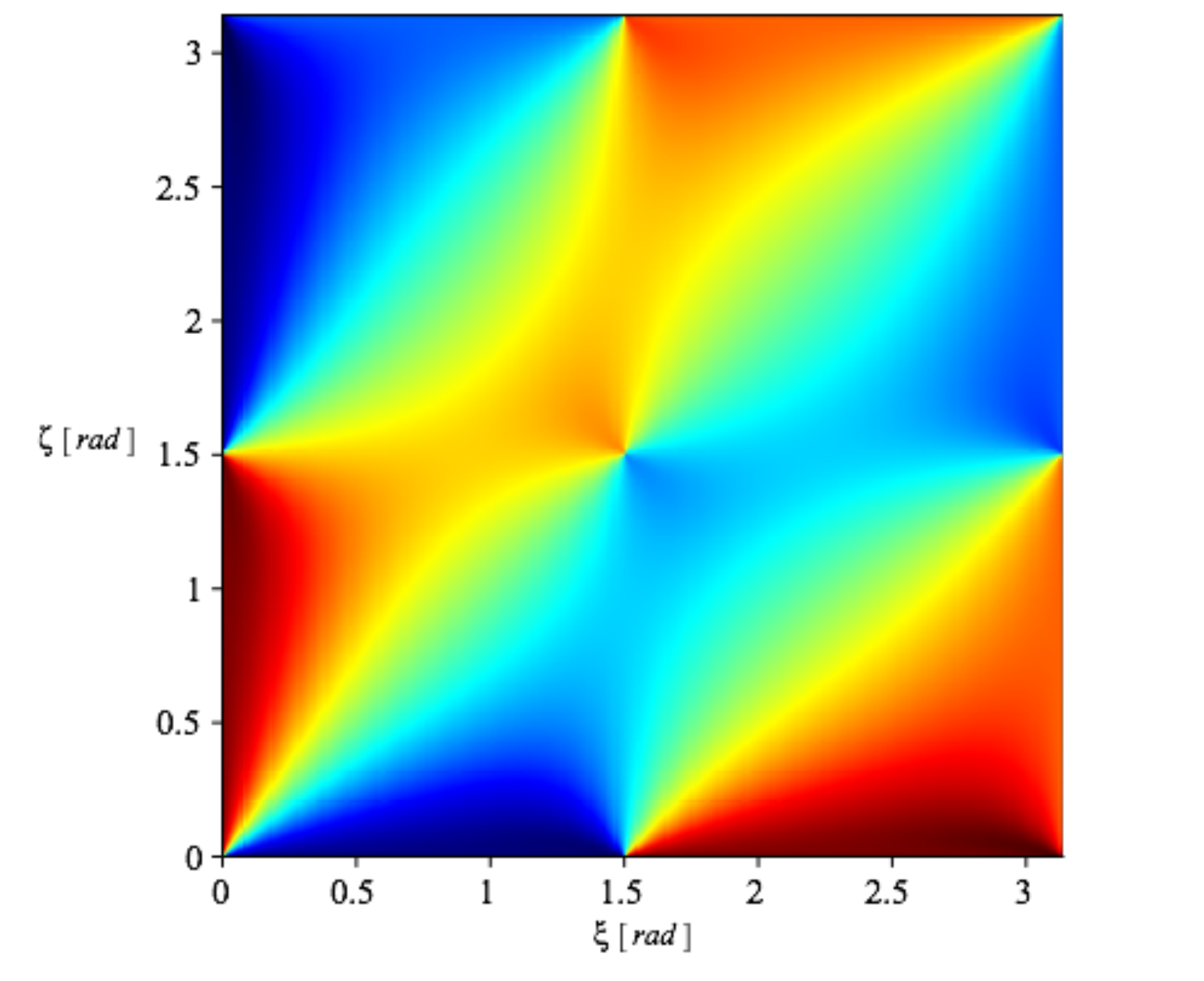}
\end{center}
\end{minipage}
&
\begin{minipage}{0.1\hsize}
\begin{center}
\includegraphics[width=9.0cm,bb=0 -45 700 400]{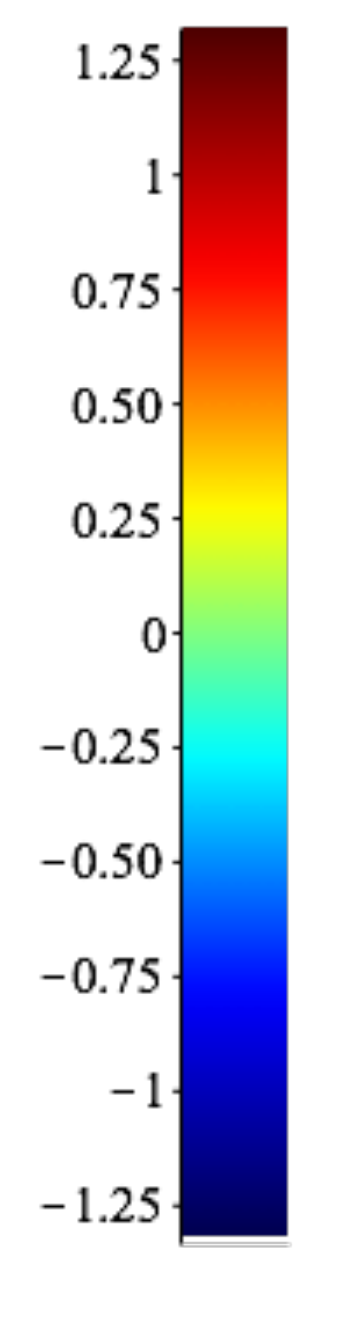}
\end{center}
\end{minipage}\\[-0.3cm]
\multicolumn{2}{c}{(a) $\Gamma_{12:34}^{I}:  \,\,( l, m)=(0,0), ( l', m')=(1,1)$}
\end{tabular}
\begin{tabular}{cc}
\begin{minipage}{0.31\hsize}
\begin{center}
\includegraphics[width=9.0cm,bb=0 0 700 400]{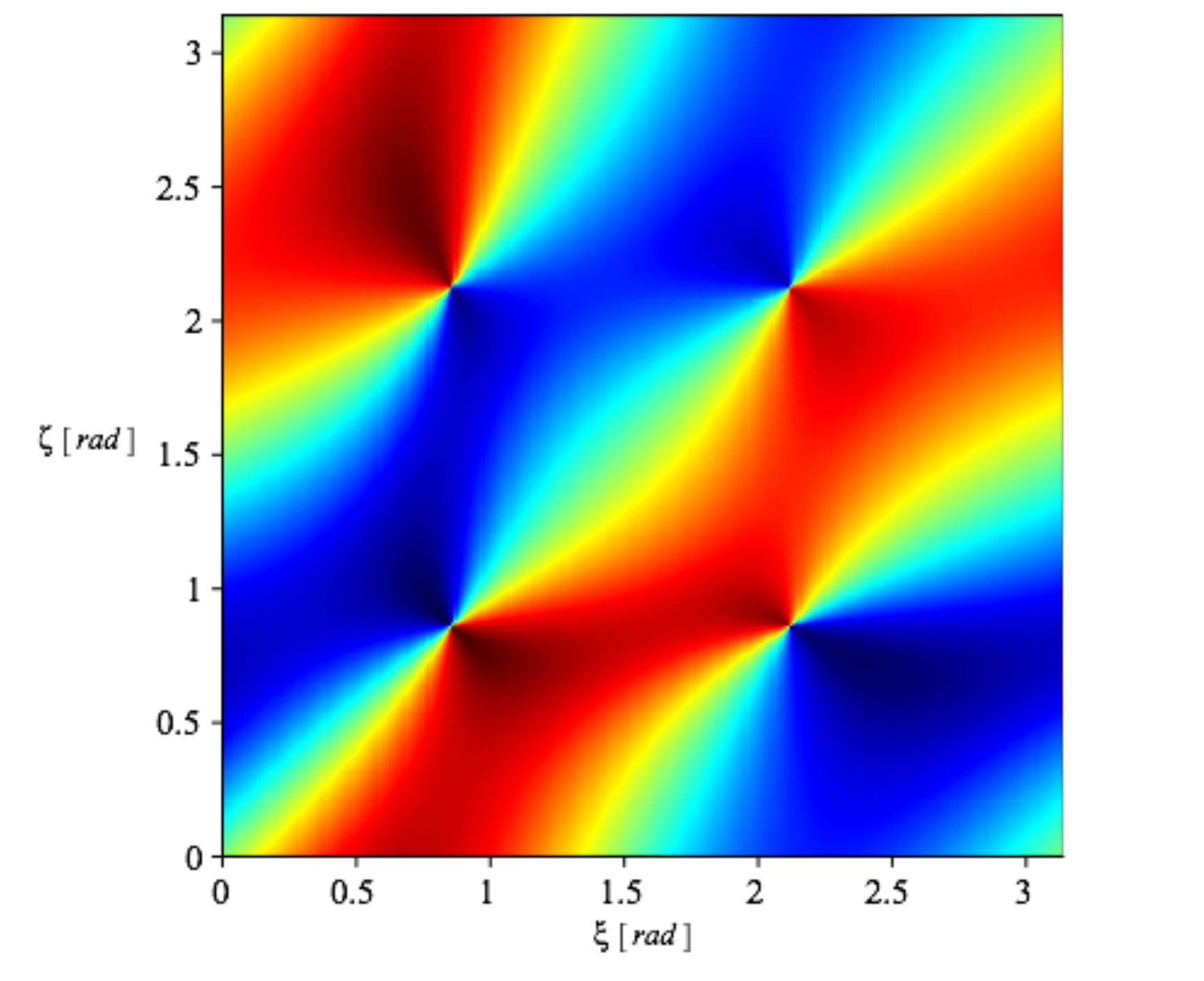}
\end{center}
\end{minipage}
&
\begin{minipage}{0.1\hsize}
\begin{center}
\includegraphics[width=9.0cm,bb=0 -45 700 400]{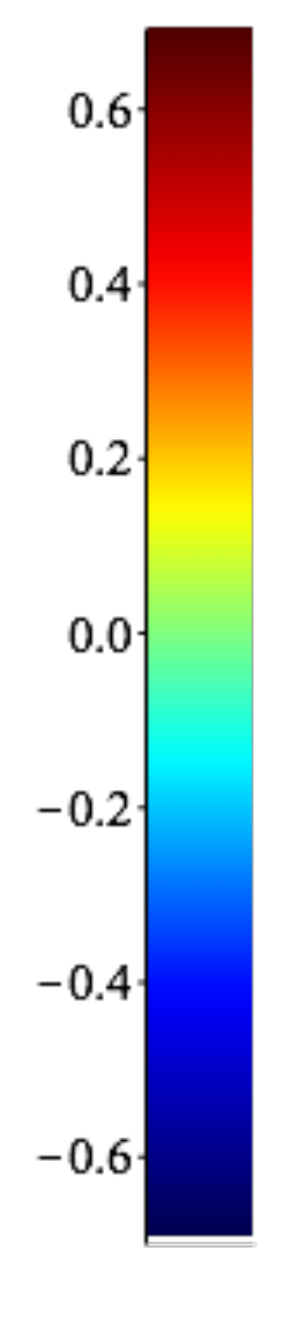}
\end{center}
\end{minipage}\\[-0.3cm]
\multicolumn{2}{c}{(b) $\Gamma_{12:34}^{V}:  \,\,( l, m)=(0,0), ( l', m')=(1,1)$}
\end{tabular}
\begin{tabular}{cc}
\begin{minipage}{0.31\hsize}
\begin{center}
\includegraphics[width=9.0cm,bb=0 0 700 400]{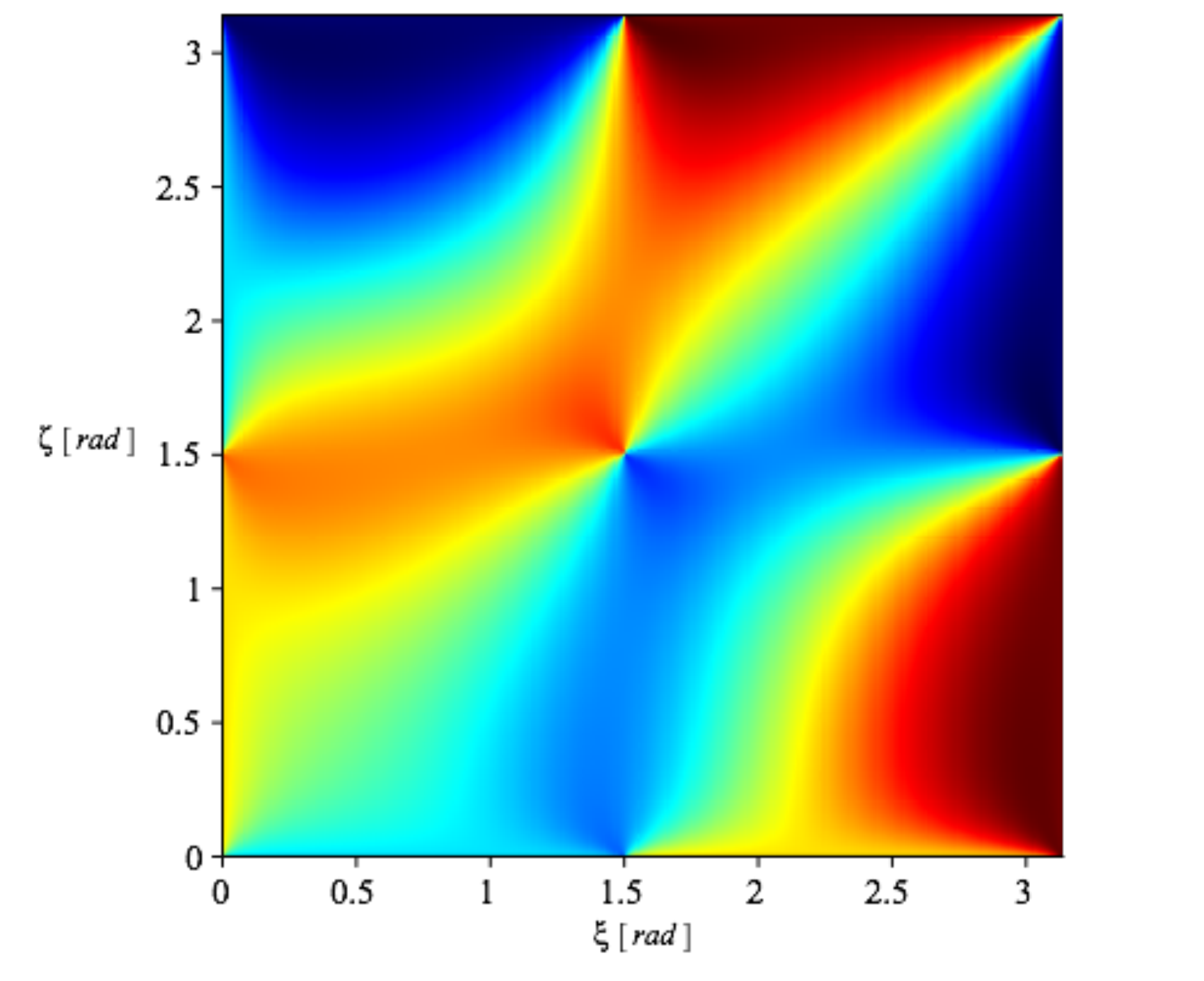}
\end{center}
\end{minipage}
&
\begin{minipage}{0.1\hsize}
\begin{center}
\includegraphics[width=9.0cm,bb=0 -45 700 400]{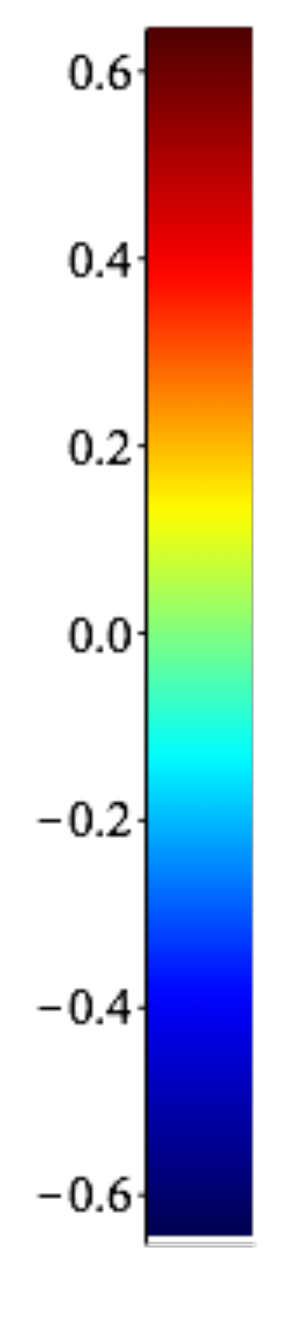}
\end{center}
\end{minipage}\\[-0.3cm]
\multicolumn{2}{c}{(c) $\Gamma_{12:34}^{I}:  \,\,( l, m)=(0,0),(1,0), ( l', m')=(1,1)$}
\end{tabular}
\begin{tabular}{cc}
\begin{minipage}{0.31\hsize}
\begin{center}
\includegraphics[width=9.0cm,bb=0 0 700 400]{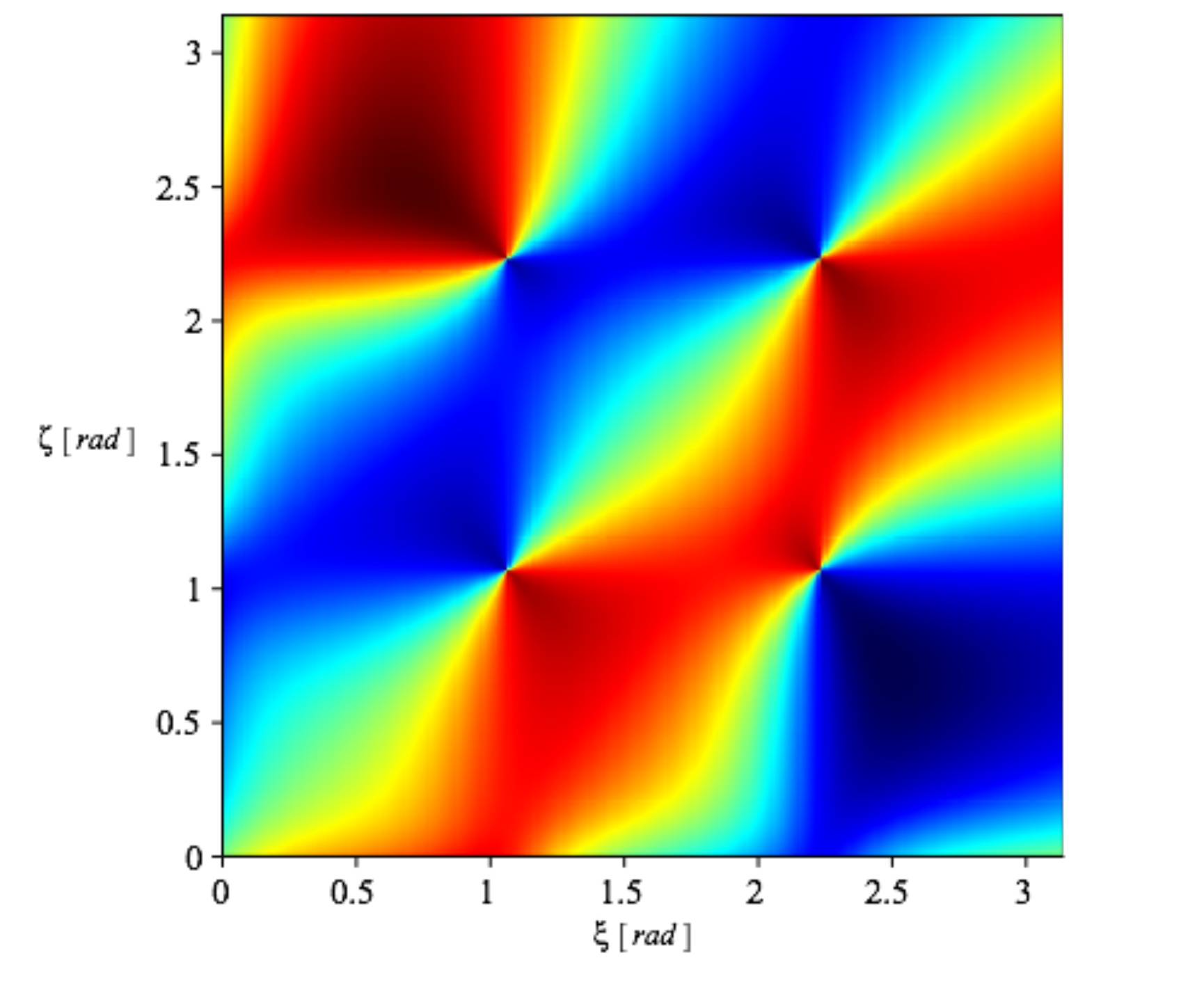}
\end{center}
\end{minipage}
&
\begin{minipage}{0.1\hsize}
\begin{center}
\includegraphics[width=9.0cm,bb=0 -45 700 400]{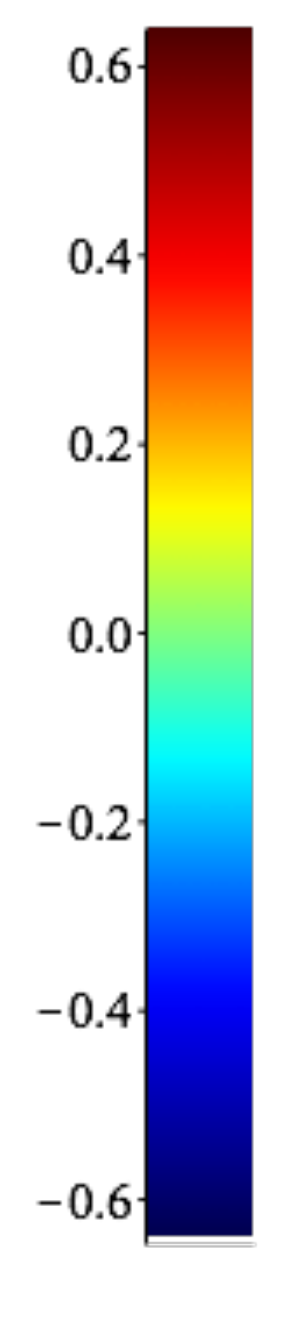}
\end{center}
\end{minipage}\\[-0.3cm]
\multicolumn{2}{c}{(d) $\Gamma_{12:34}^{V}:  \,\,( l, m)=(0,0),(1,0), ( l', m')=(1,1)$}
\end{tabular}
\begin{tabular}{cc}
\begin{minipage}{0.31\hsize}
\begin{center}
\includegraphics[width=9.0cm,bb=0 0 700 400]{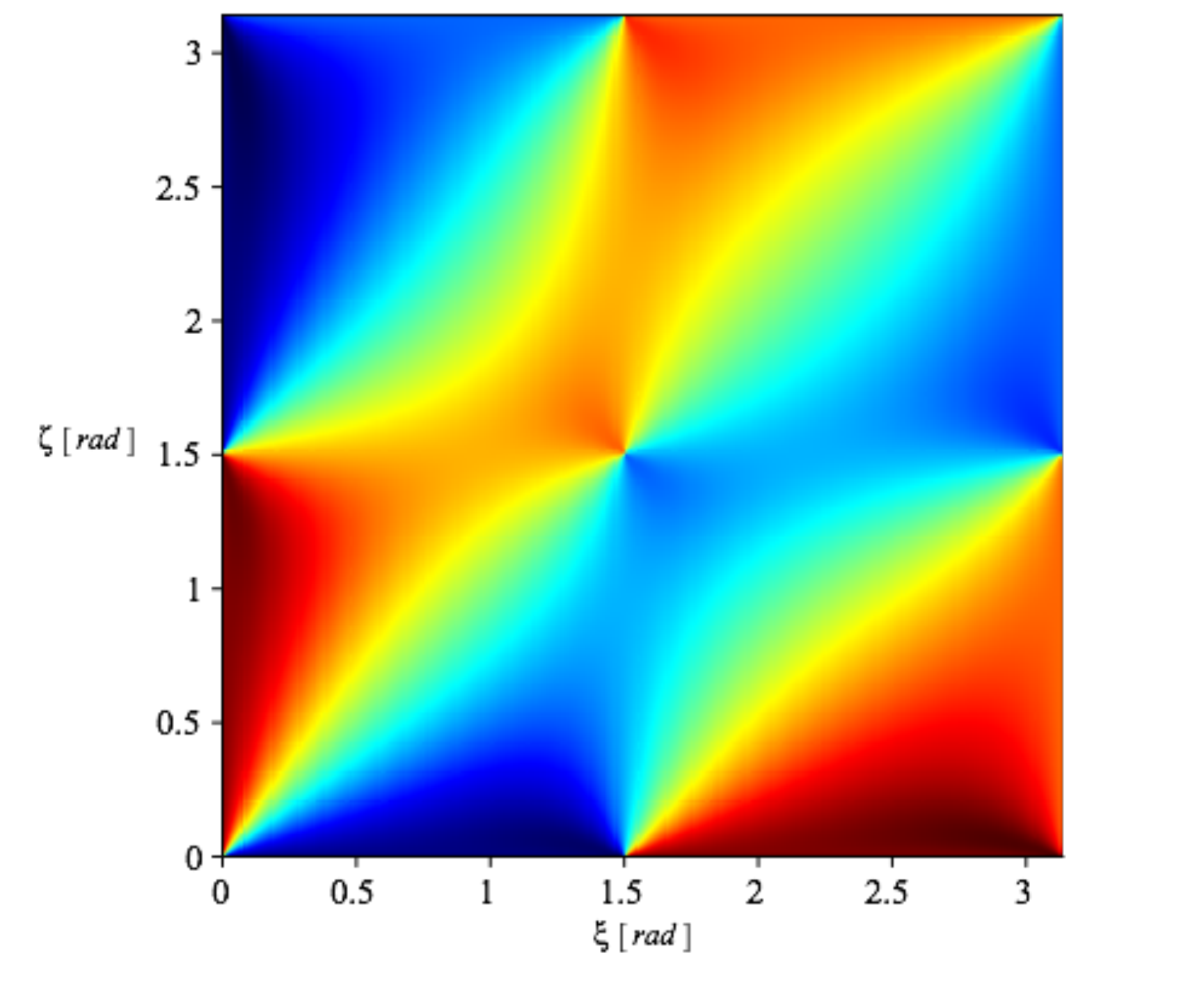}
\end{center}
\end{minipage}
&
\begin{minipage}{0.1\hsize}
\begin{center}
\includegraphics[width=9.0cm,bb=0 -45 700 400]{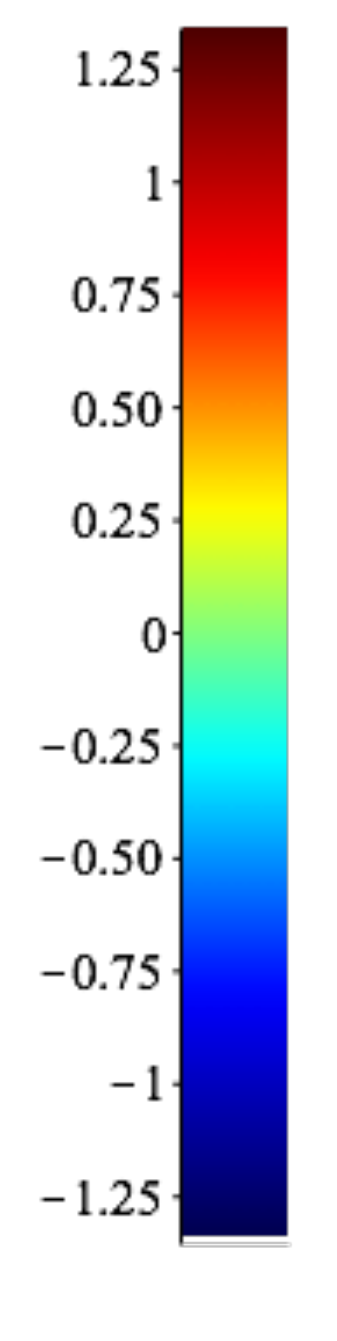}
\end{center}
\end{minipage}\\[-0.3cm]
\multicolumn{2}{c}{(e) $\Gamma_{12:34}^{I}:  \,\,( l, m)=(0,0),(1,1), ( l', m')=(1,1)$}
\end{tabular}
\begin{tabular}{cc}
\begin{minipage}{0.31\hsize}
\begin{center}
\includegraphics[width=9.0cm,bb=0 0 700 400]{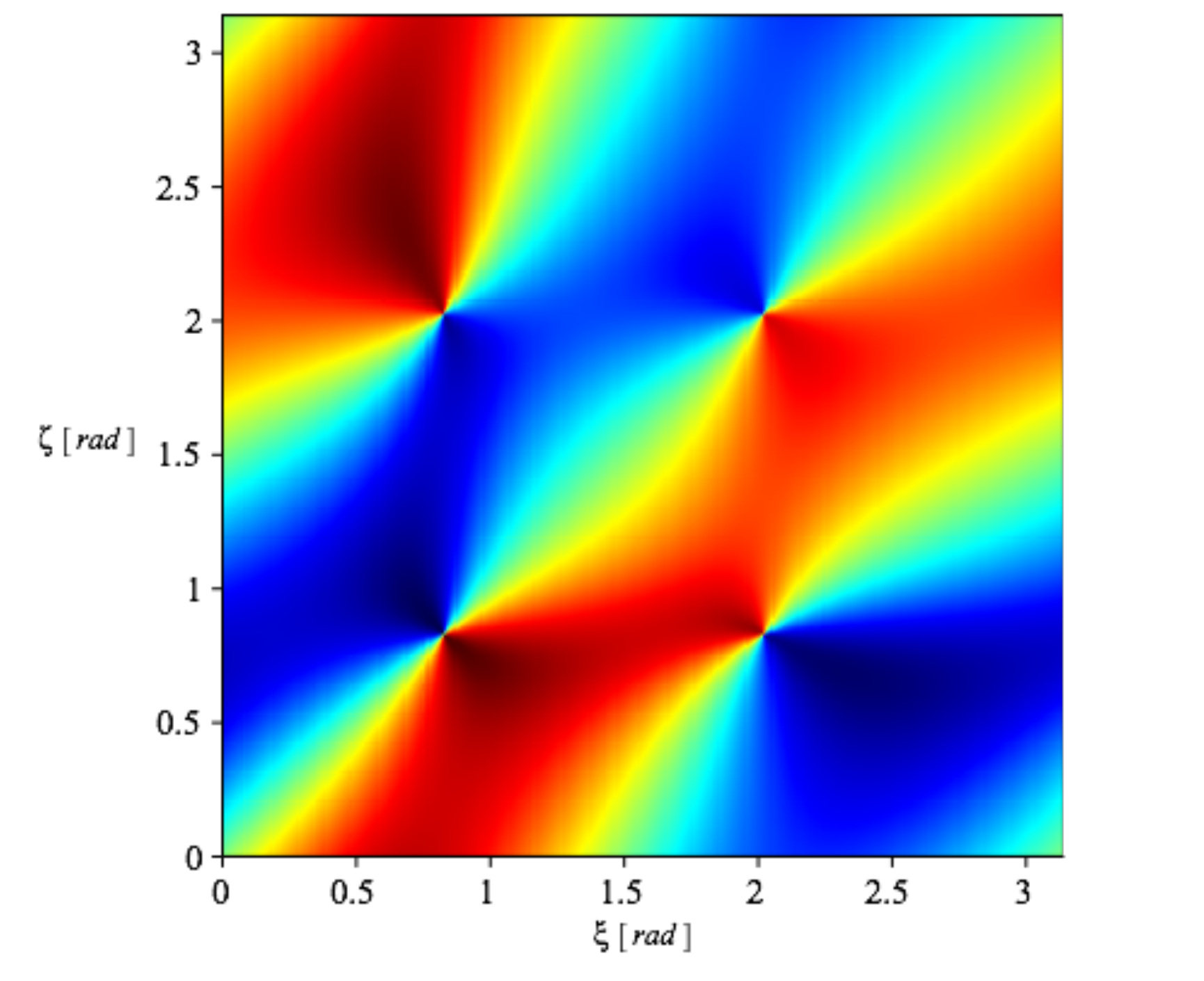}
\end{center}
\end{minipage}
&
\begin{minipage}{0.1\hsize}
\begin{center}
\includegraphics[width=9.0cm,bb=0 -45 700 400]{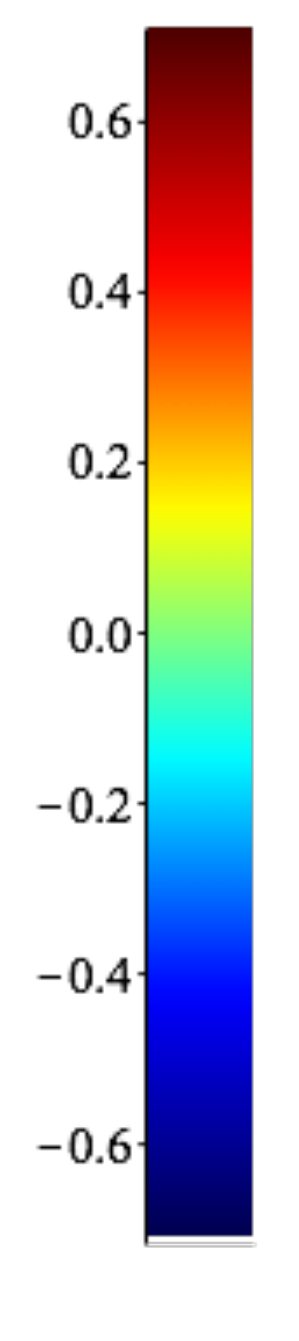}
\end{center}
\end{minipage}\\[-0.3cm]
\multicolumn{2}{c}{(f) $\Gamma_{12:34}^{V}:  \,\,( l, m)=(0,0),(1,1), ( l', m')=(1,1)$}
\end{tabular}
\begin{tabular}{cc}
\begin{minipage}{0.31\hsize}
\begin{center}
\includegraphics[width=9.0cm,bb=0 0 700 400]{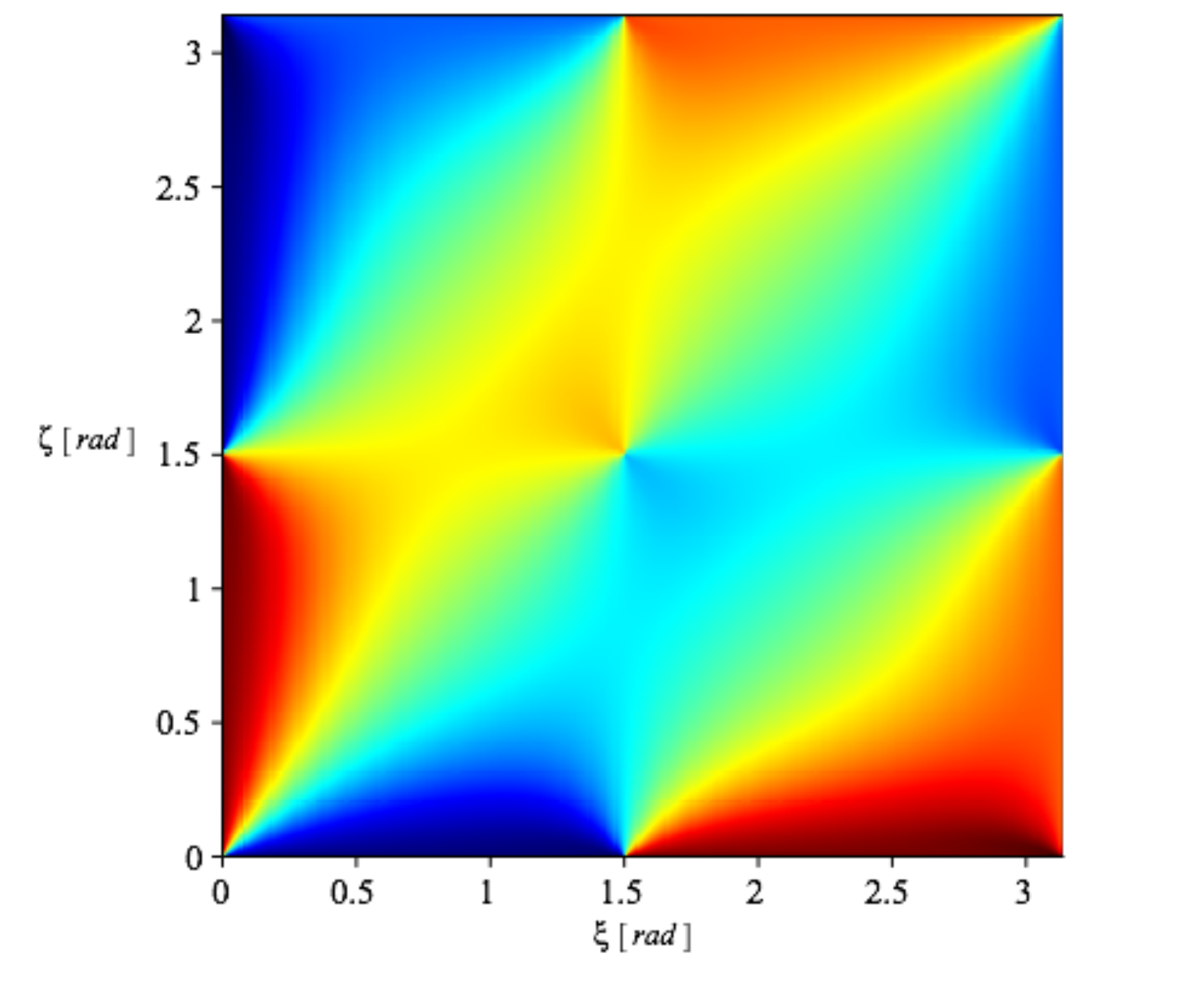}
\end{center}
\end{minipage}
&
\begin{minipage}{0.1\hsize}
\begin{center}
\includegraphics[width=9.0cm,bb=0 -45 700 400]{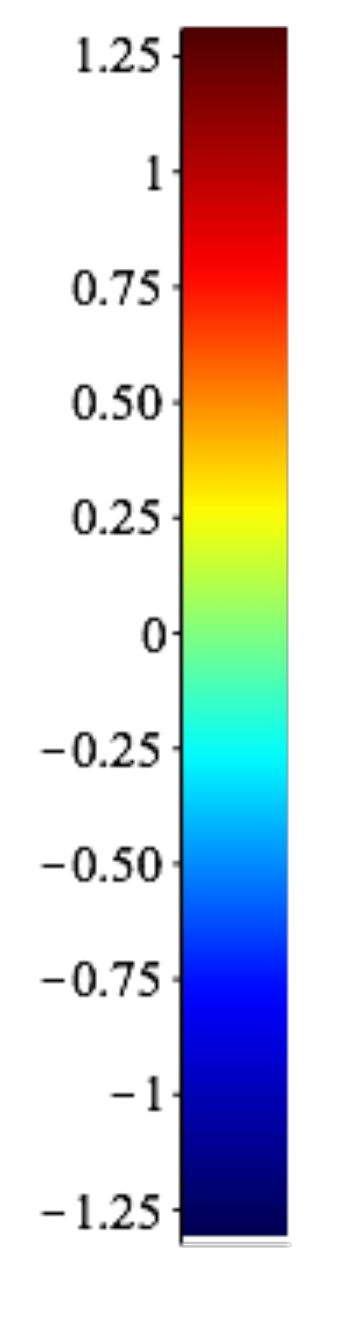}
\end{center}
\end{minipage}\\[-0.3cm]
\multicolumn{2}{c}{(g) $\Gamma_{12:34}^{I}:  \,\,( l, m)=(0,0),(1,-1), ( l', m')=(1,1)$}
\end{tabular}
\begin{tabular}{cc}
\begin{minipage}{0.31\hsize}
\begin{center}
\includegraphics[width=9.0cm,bb=0 0 700 400]{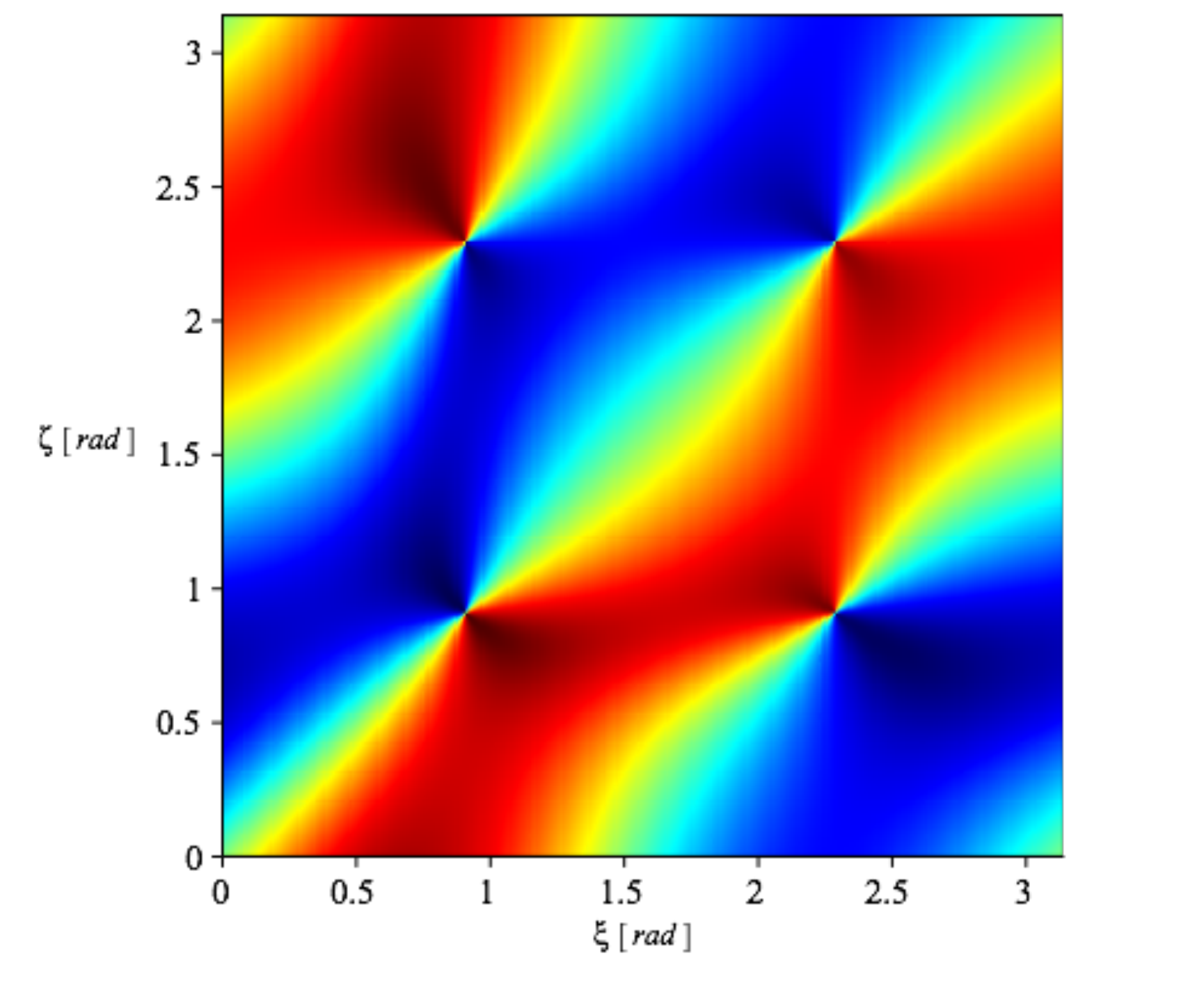}
\end{center}
\end{minipage}
&
\begin{minipage}{0.1\hsize}
\begin{center}
\includegraphics[width=9.0cm,bb=0 -45 700 400]{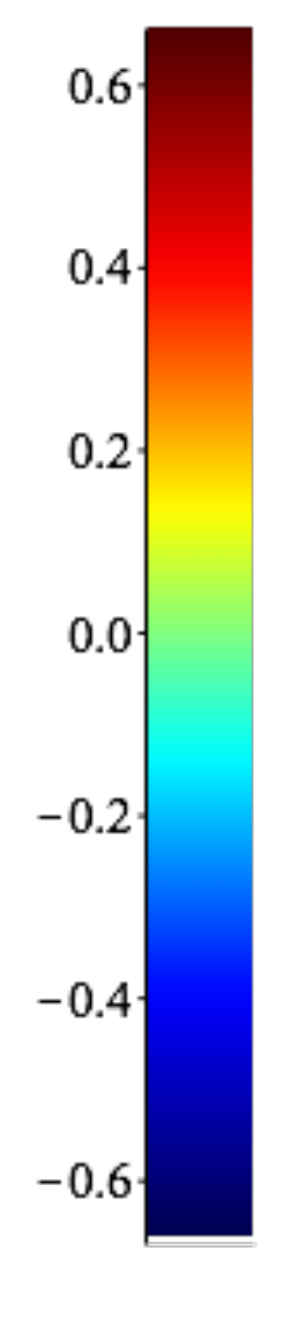}
\end{center}
\end{minipage}\\[-0.3cm]
\multicolumn{2}{c}{(h) $\Gamma_{12:34}^{V}:  \,\,( l, m)=(0,0),(1,-1), ( l', m')=(1,1)$}
\end{tabular}
\caption{Density plots of the compiled ORFs  $\Gamma_{12:34}^{I}$ (left panels) and  $\Gamma_{12:34}^{V}$ (right panels) as a function of the two angular separations $\xi,\zeta$ for two pulsar pairs, respectively.
For simplicity, we assumed $c^{I}_{10}/c^{I}_{00}=c^{I}_{11}/c^{I}_{00}=c^{I}_{1-1}/c^{I}_{00}=1$.}
\label{CG}
\end{figure}

In Fig. \ref{CG} we show some compiled ORFs  $\Gamma_{12:34}^{I}$ (left panels) and $\Gamma_{12:34}^{V}$ (right panels) as a function of the two angular separations $\xi$ and $\zeta$ for two pulsar pairs, respectively.
We used the expressions of $V$ mode and $I$ mode (see Appendix \ref{sec:the generalized overlap reduction function for intensity}), and  we assumed $c^{I}_{10}/c^{I}_{00}=c^{I}_{11}/c^{I}_{00}=c^{I}_{1-1}/c^{I}_{00}=1$ for simplicity.
In Fig. \ref{CG}(a) and \ref{CG}(b),  the $I$ mode is dominated  by $(l, m)=(0,0)$ and $V$ mode is dominated by $(l', m')=(1,1)$. 
In Fig. \ref{CG}(c) and \ref{CG}(d),  the $I$ mode is dominated by $(l, m)=(0,0),(1,0)$ and $V$ mode is dominated by $(l', m')=(1,1)$.
In Fig. \ref{CG}(e) and \ref{CG}(f),  the $I$ mode is dominated by $(l, m)=(0,0),(1,1)$ and $V$ mode is dominated by $(l', m')=(1,1)$.
In Fig. \ref{CG}(e) and \ref{CG}(f),  the $I$ mode is dominated by $(l, m)=(0,0),(1,-1)$ and $V$ mode is dominated by $(l', m')=(1,1)$.
By definition, in the case of $\xi=\zeta$, the compiled ORFs are zero.


\section{Conclusion}
We have studied the detectability of the stochastic gravitational waves with PTAs. 
In most of the previous works, the isotropy of SGWB has been assumed for the analysis.
Recently,  however, a stochastic gravitational wave background with anisotropy have been considered.
The information of the anisotropic pattern of the distribution should contain important information of the sources such as 
supermassive black hole binaries and the sources in the early universe. It is also intriguing to take into account the polarization of SGWB
in the PTA analysis.   
Therefore, we extended the correlation analysis to circularly  polarized SGWB
 and calculated generalized overlap reduction functions for them. It turned out  that the circular polarization can not be detected 
for an isotropic background. However, when the distribution has anisotropy, we have shown that
 there is a chance to observe circular polarizations in the SGWB.

We also discussed how to separate polarized modes from unpolarized modes of gravitational waves. 
If we have a priori knowledge of the abundance ratio for each mode in each of $l,m$, we can separate $I$ mode and $V$ mode in general.
This would be possible if we start from fundamental theory and calculate the spectrum of SGWB. 
In particular, in the case that the signal of lowest $(l,m)$ is dominant,  we performed the  separation of $I$ mode and $V$ mode explicitly.

\acknowledgments
This work was supported by  Grants-in-Aid for Scientific Research (C) No.25400251 and "MEXT Grant-in-Aid for Scientific Research on Innovative Areas  No.26104708 and “Cosmic Acceleration”(No.15H05895) .

\appendix

\section{generalized overlap reduction function for circular polarization: dipole cases}\label{sec:angular integral of the generalized overlap reduction function for dipole circular polarization}
In this Appendix, we perform angular integration of the generalized ORF for dipole ($l=1$) circular polarization (see \cite{Mingarelli:2013dsa,Jenet:2014bea}):
\begin{eqnarray}
\Gamma^{V}_{1m}&=&-\frac{N_{1}^{m}}{2}\int_{0}^{\pi}d\theta\,\,\sin\theta (1-\cos\theta)P_{1}^{m}(\cos\theta)\int_{0}^{2\pi}d\phi\,\,\frac{(\sin^2\xi\cos\theta\sin\phi\cos\phi-\sin\xi\cos\xi\sin\theta\sin\phi)}{1+\sin\xi\sin\theta\cos\phi+\cos\xi\cos\theta}\sin(m\phi)
\nonumber\\
&=&-\frac{N_{1}^{m}}{2}\int_{-1}^{1}dx\,\, (1-x)P_{1}^{m}(x)\int_{0}^{2\pi}d\phi\,\,\frac{(x\sin^2\xi\sin\phi\cos\phi-\sqrt{1-x^2}\sin\xi\cos\xi\sin\phi)}{1+\sqrt{1-x^2}\sin\xi\cos\phi+x\cos\xi}\sin(m\phi)\ ,\label{g1m}
\end{eqnarray}
where we have defined $x\equiv\cos\theta$. It is obvious that in the case of $(l,m)=(1,0)$, integrand of the generalized ORF is zero, because of $\sin(0)=0$, then we obtain
\begin{eqnarray}
\Gamma^{V}_{10}&=&0\ .
\end{eqnarray}
Then, using \erefs{nlm}-\eqref{pl}, we calculate
\begin{eqnarray}
N_{1}^{1} = \sqrt{\frac{3}{8\pi}} \ , \quad
N_{1}^{-1} = \sqrt{\frac{3}{2\pi}} \ , \quad
P_{1}^{1}(x) = -\sqrt{1-x^{2}} \  , \quad
P_{1}^{-1}(x) = \frac{\sqrt{1-x^{2}}}{2} \ ,
\end{eqnarray}
and we find
\begin{eqnarray}
\Gamma^{V}_{11}=\Gamma^{V}_{1-1} \ .
\end{eqnarray}

Therefore we only have to consider the dipole generalized ORF in the case of $l=1$, $m=1$:
\begin{eqnarray}
\Gamma^{V}_{11}&=&\frac{1}{2}\sqrt{\frac{3}{8\pi}}\int_{-1}^{1}dx\,\, (1-x)\sqrt{1-x^{2}}\int_{0}^{2\pi}d\phi\,\,\frac{(x\sin^2\xi\sin\phi\cos\phi-\sqrt{1-x^2}\sin\xi\cos\xi\sin\phi)}{1+\sqrt{1-x^2}\sin\xi\cos\phi+x\cos\xi}\sin\phi \nonumber\\
&=&\frac{1}{2}\sqrt{\frac{3}{8\pi}}(G(\xi)+L(\xi))\ ,\label{gal}
\end{eqnarray}
where
\begin{eqnarray}
G(\xi)&\equiv&\sin^2\xi\int_{-1}^{1}dx\,\, x(1-x)\sqrt{1-x^{2}}H(x,\xi)\ ,\label{a}\\
H(x,\xi)&\equiv&\int_{0}^{2\pi}d\phi\,\,\frac{\sin^{2}\phi\cos\phi}{1+\sqrt{1-x^2}\sin\xi\cos\phi+x\cos\xi}\ ,\label{b}\\
L(\xi)&\equiv&-\sin\xi\cos\xi\int_{-1}^{1}dx\,\, (1-x)(1-x^2)M(x,\xi)\ ,\label{l}\\
M(x,\xi)&\equiv&\int_{0}^{2\pi}d\phi\,\,\frac{\sin^{2}\phi}{1+\sqrt{1-x^2}\sin\xi\cos\phi+x\cos\xi}\ .\label{m}
\end{eqnarray}
First, to calculate $M(x,\xi)$, we use contour integral in the complex plane.
Defining $z=e^{i\phi}$ and substituting 
\begin{eqnarray}
\cos\phi=\frac{z+z^{-1}}{2} \ ,\quad \sin\phi=\frac{z-z^{-1}}{2i} \ ,\quad  d\phi=\frac{dz}{iz}\ , \label{cpsp}
\end{eqnarray}
into \eref{m}, we can rewrite $M(x,\xi)$ as
\begin{eqnarray}
M(x,\xi)=\oint_{C}dz\,\,\frac{i(z^2-1)^2}{z^2[4z(1+x\cos\xi)+2\sqrt{1-x^2}\sin\xi(z^2+1)]} \ ,
\end{eqnarray}
where $C$ denotes a unit circle.
We can factorize the denominator of the integrand and  get
\begin{eqnarray}
M(x,\xi)=\oint_{C}dz\,\,\frac{i(z^2-1)^2}{2\sqrt{1-x^2}\sin\xi \cdot z^{2}(z-z_{+})(z-z_{-})} \ ,
\end{eqnarray}
where
\begin{eqnarray}
z_{+}\equiv-\displaystyle\sqrt{\frac{(1\mp x)(1\mp\cos\xi)}{(1\pm x)(1\pm\cos\xi)}}\ ,\quad z_{-}\equiv\frac{1}{z_{+}} \ .
\end{eqnarray}
Hereafter, the upper sign applies when $-\cos\xi\le x\le1$ and the lower one applies when $-1\le x\le-\cos\xi$.
Note that we only consider the region $0\le\xi\le\pi$, so we have used the relation $\sin\xi=\sqrt{1-\cos^2\xi}$ in above expression.
In the region $-1\le x\le 1$, $z_{+}$ is inside the unit circle $C$ except for $x=-\cos\xi$ and $z_{-}$ is outside the unit circle $C$.
Now, we can perform the integral using the residue theorem
\begin{eqnarray}
M(x,\xi)=\oint_{C}dz\,\,f(z)=2\pi i\sum_{i}{\rm Res}(f,z_{i})\ ,
\end{eqnarray}
where
\begin{eqnarray}
f(z)\equiv\frac{i(z^2-1)^2}{2\sqrt{1-x^2}\sin\xi \cdot z^{2}(z-z_{+})(z-z_{-})} \ .
\end{eqnarray}
The residues inside the unit circle $C$ can be evaluated as
\begin{eqnarray}
{\rm Res}(f,0) = \lim_{z\rightarrow 0}\left\{\frac{d}{dz}[z^2f(z)]\right\}
           = \frac{i(z_{+}+z_{-})}{2\sqrt{1-x^2}\sin\xi} \ ,
\end{eqnarray}
\begin{eqnarray}
{\rm Res}(f,z_{+}) = \lim_{z\rightarrow z_{+}}\{(z-z_{+})f(z)\}
     = \frac{i(z_{+}-z_{-})}{2\sqrt{1-x^2}\sin\xi} \ .
\end{eqnarray}
Thus, we obtain
\begin{eqnarray}
M(x,\xi)=\frac{2\pi}{(1\pm x)(1\pm\cos\xi)}\ .\label{m2}
\end{eqnarray}
Next, we consider $L(\xi)$ defined in \eref{l}.
Using \eref{m2}, we can calculate $L(\xi)$ as
\begin{eqnarray}
L(\xi)&=&-2\pi\sin\xi\cos\xi\int_{-1}^{1}dx\,\,\frac{(1-x^2)(1-x)}{(1\pm x)(1\pm\cos\xi)}\nonumber\\
&=&-2\pi\sin\xi\cos\xi\left\{\frac{1}{(1-\cos\xi)}\int_{-1}^{-\cos\xi}dx\,\,(1-x^2)+\frac{1}{(1+\cos\xi)}\int_{-\cos\xi}^{1}dx\,\,(1-x)^2\right\}  \nonumber\\
&=&-2\pi\sin\xi\cos\xi\left(1+\frac{1}{3}\cos\xi\right)\ .\label{l2}
\end{eqnarray}

Similarly, we can evaluate  $H(x,\xi)$ given in \eref{b}.
To calculate $H(x,\xi)$ in the complex plane, we again substitute \eref{cpsp} into \eref{b} and obtain
\begin{eqnarray}
H(x,\xi)=\oint_{C}dz\,\,\frac{i(z^2-1)^2(z^2+1)}{4\sqrt{1-x^2}\sin\xi \cdot z^{3}(z-z_{+})(z-z_{-})} \ .
\end{eqnarray}
We use the residue theorem
\begin{eqnarray}
H(x,\xi)=\oint_{C}dz\,\,g(z)=2\pi i\sum_{i}{\rm Res}(g,z_{i})\ ,
\end{eqnarray}
where
\begin{eqnarray}
g(z)&\equiv&\frac{i(z^2-1)^2(z^2+1)}{4\sqrt{1-x^2}\sin\xi \cdot z^{3}(z-z_{+})(z-z_{-})} \ .
\end{eqnarray}
The residues inside the unit circle $C$ can be calculated as
\begin{eqnarray}
{\rm Res}(g,0) = \lim_{z\rightarrow 0}\left\{\frac{d^2}{dz^2}\left[\frac{1}{2}z^3g(z)\right]\right\}
      =\frac{i(z_{+}^2+z_{-}^2)}{4\sqrt{1-x^2}\sin\xi} \ ,
\end{eqnarray}
\begin{eqnarray}
{\rm Res}(g,z_{+}) = \lim_{z\rightarrow z_{+}}\{(z-z_{+})g(z)\}
  =  \frac{i(z^2_{+}-z^2_{-})}{4\sqrt{1-x^2}\sin\xi}\ .
\end{eqnarray}
Therefore, $H(x,\xi)$ becomes
\begin{eqnarray}
H(x,\xi)&=&\frac{-\pi}{(1\pm x)(1\pm\cos\xi)}\sqrt{\frac{(1\mp x)(1\mp\cos\xi)}{(1\pm x)(1\pm\cos\xi)}}\ .\label{b2}
\end{eqnarray}
Substituting \eref{b2} to \eref{a}, we can calculate $G(\xi)$:
\begin{eqnarray}
G(\xi)&=&-\pi\sin^2\xi\int_{-1}^{1}dx\,\,\frac{x(1-x)\sqrt{1-x^2}}{(1\pm x)(1\pm\cos\xi)}\sqrt{\frac{(1\mp x)(1\mp\cos\xi)}{(1\pm x)(1\pm\cos\xi)}} \nonumber\\
&=&-\pi\sin^2\xi\left\{\frac{1}{1-\cos\xi}\sqrt{\frac{1+\cos\xi}{1-\cos\xi}}\int_{-1}^{-\cos\xi}dx\,\,x(1+x)+\frac{1}{1+\cos\xi}\sqrt{\frac{1-\cos\xi}{1+\cos\xi}}\int_{-\cos\xi}^{1}dx\,\,\frac{x(1-x)^2}{1+x}\right\}  \nonumber\\
&=&-\frac{2\pi}{3}\sin\xi\left\{4-3\cos\xi-\cos^2\xi+\left(\frac{1-\cos\xi}{1+\cos\xi}\right)\left(-6\log2+6\log(1-\cos\xi)\right)\right\}.\label{a2}
\end{eqnarray}

Finally, substituting \erefs{l2} and \eqref{a2} into \eref{gal}, we get the generalized ORF for $(l,m)=(1,1)$ 
\begin{eqnarray}
\Gamma^{V}_{11}&=&-\frac{\sqrt{6\pi}}{3}\sin\xi\left[1+3\left(\frac{1-\cos\xi}{1+\cos\xi}\right)\log\left(\sin\frac{\xi}{2}\right)\right]\ .
\end{eqnarray}

\begin{figure}[t]
\begin{minipage}{0.4\hsize}
\begin{center}
\includegraphics[width=9.0cm,bb=0 0 700 400]{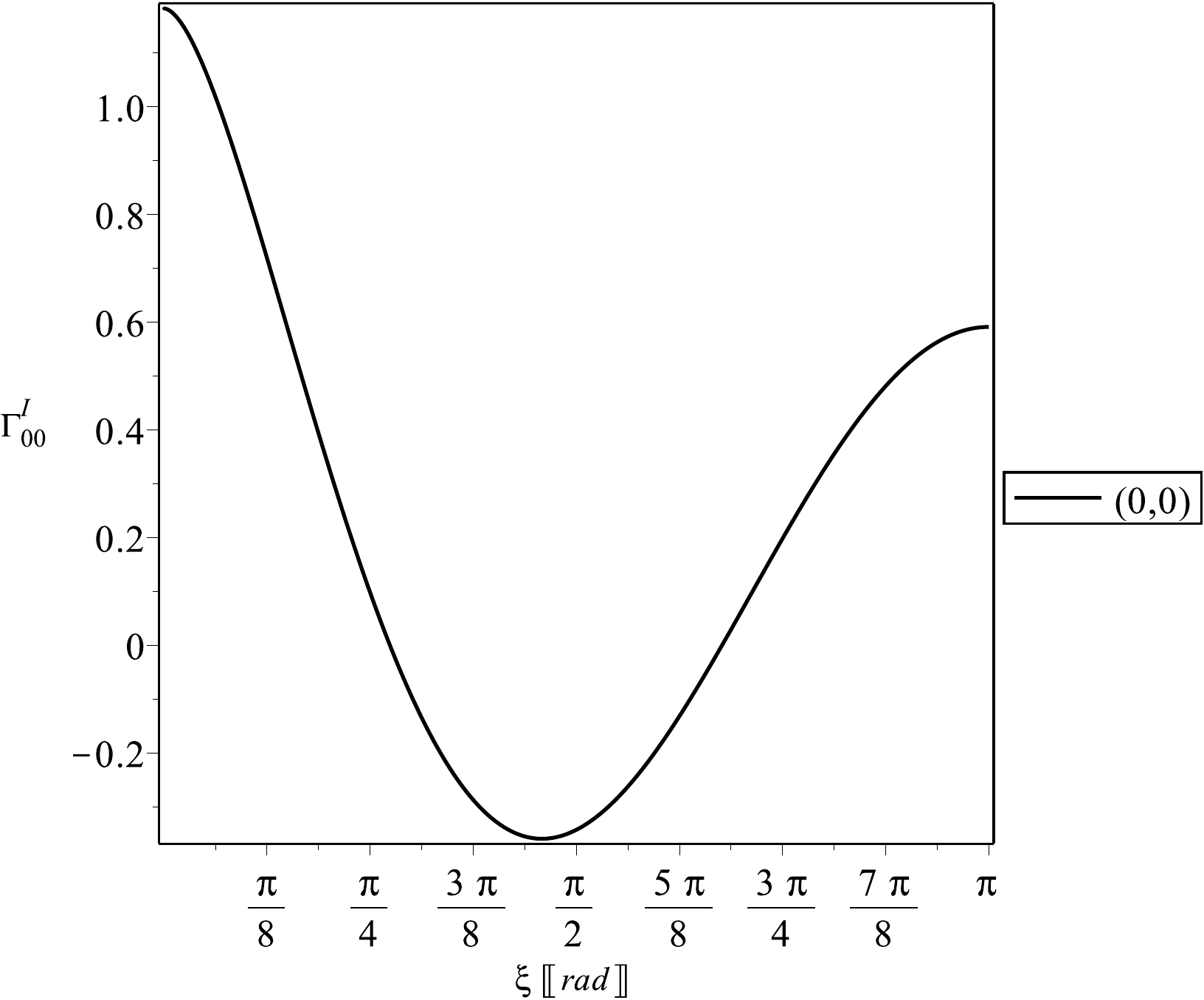}
\hspace{1.6cm} (a) $l=0$
\end{center}
\end{minipage}
\begin{minipage}{0.4\hsize}
\begin{center}
\includegraphics[width=9.0cm,bb=0 0 700 400]{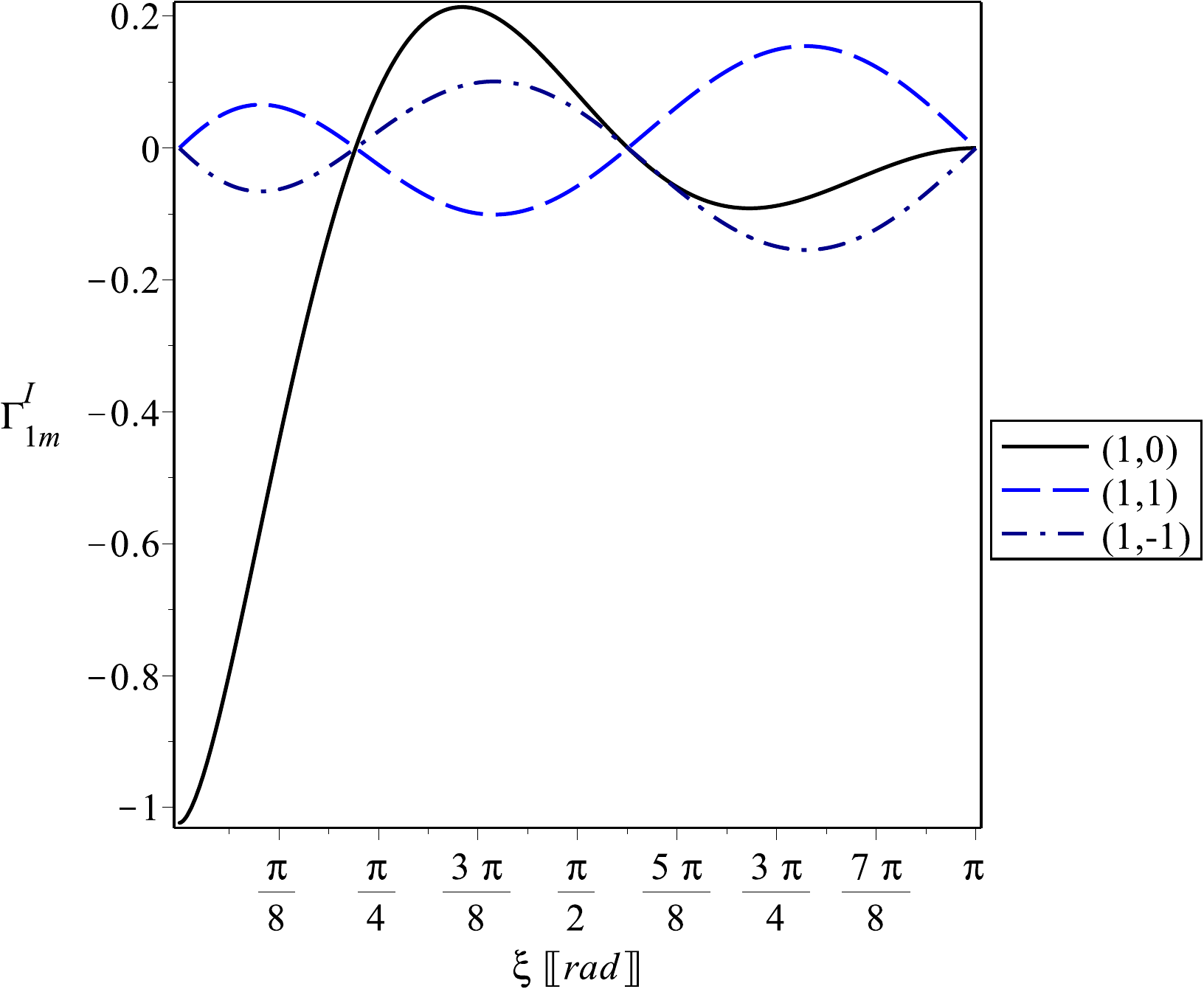}
\hspace{1.6cm} (b) $l=1$
\end{center}
\end{minipage}
\begin{minipage}{0.4\hsize}
\begin{center}
\includegraphics[width=9.0cm,bb=0 0 700 400]{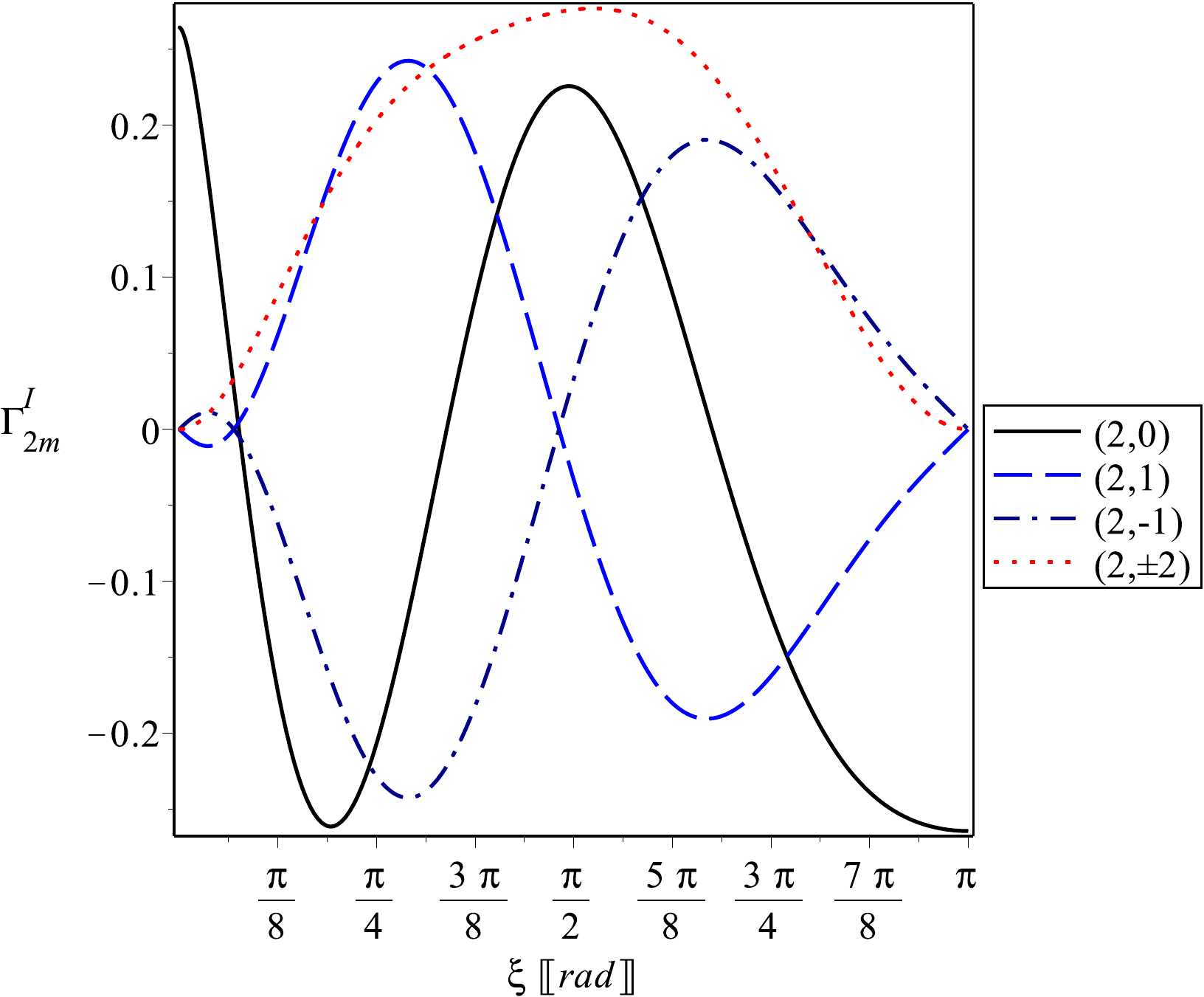}
\hspace{1.6cm} (c) $l=2$
\end{center}
\end{minipage}
\begin{minipage}{0.4\hsize}
\begin{center}
\includegraphics[width=9.0cm,bb=0 0 700 400]{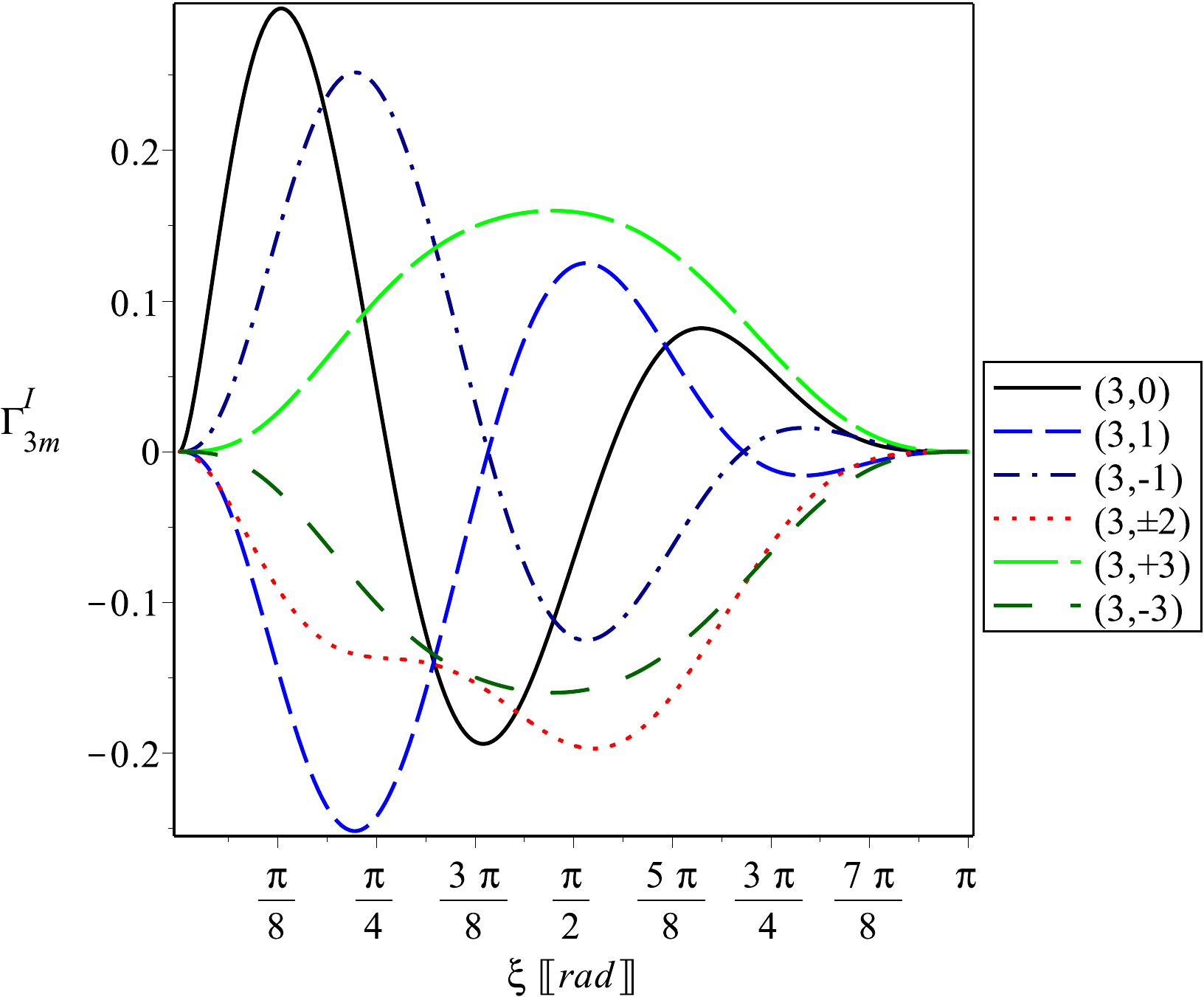}
\hspace{1.6cm} (d) $l=3$
\end{center}
\end{minipage}
\caption{Plots  the generalized ORFs $\Gamma_{lm}^{I}$ as a function of the angular separation between the two pulsars $\xi$.
Fig. \ref{GI}(a) shows monopole (l=0), Fig. \ref{GI}(b) shows dipole (l=1), Fig. \ref{GI}(c) shows quadrupole (l=2) and Fig. \ref{GI}(d) shows octupole (l=3).
The black solid curve, the blue dashed curve, the dark-blue dash-dotted curve, the red dotted curve, the green long-dashed curve, 
 the dark-green space-dashed curve represent $m=0$, $m=+1$, $m=-1$, $m=\pm 2$, $m=+3$, $m=-3$, respectively.}
\label{GI}
\end{figure}

\section{Generalized overlap reduction function for intensity}\label{sec:the generalized overlap reduction function for intensity}
In this Appendix,we show ORFs for the intensity \cite{Mingarelli:2013dsa}.
The following form for  $l=0$ was derived in \cite{Hellings:1983fr}, 
and our expressions are identical to their expressions:
\begin{eqnarray}
\Gamma^{I}_{00}&=&\frac{\sqrt{\pi}}{2}\left[1+\frac{\cos\xi}{3}+4(1-\cos\xi)\log\left(\sin\frac{\xi}{2}\right)\right]\ ,
\end{eqnarray}
For, $l=1$, we calculated as
\begin{eqnarray}
\Gamma^{I}_{10}&=&-\frac{\sqrt{3\pi}}{6}\left[1+\cos\xi+3(1-\cos\xi)\left\{1+\cos\xi+4\log\left(\sin\frac{\xi}{2}\right)\right\}\right]\ ,\\
\Gamma^{I}_{11}&=&\frac{\sqrt{6\pi}}{12}\sin\xi\left[1+3(1-\cos\xi)\left\{1+\frac{4}{1+\cos\xi}\log\left(\sin\frac{\xi}{2}\right)\right\}\right]\ ,\\
\Gamma^{I}_{1-1}&=&-\Gamma^{I}_{11}\ ,
\end{eqnarray}
For $l=2$, we obtain
\begin{eqnarray}
\Gamma^{I}_{20}&=&\frac{\sqrt{5\pi}}{60}\left[45+19\cos\xi-45\cos^2\xi-15\cos^3\xi+120(1-\cos\xi)\log\left(\sin\frac{\xi}{2}\right)\right]\ ,\\
\Gamma^{I}_{21}&=&-\frac{\sqrt{30\pi}}{60}\sin\xi\left[21-15\cos\xi-5\cos^2\xi+60\left(\frac{1-\cos\xi}{1+\cos\xi}\right)\log\left(\sin\frac{\xi}{2}\right)\right]\ ,\\
\Gamma^{I}_{2-1}&=&-\Gamma^{I}_{2-1}\ ,\\
\Gamma^{I}_{22}&=&\frac{\sqrt{30\pi}}{24}(1-\cos\xi)\left[9-4\cos\xi-\cos^2\xi+24\left(\frac{1-\cos\xi}{1+\cos\xi}\right)\log\left(\sin\frac{\xi}{2}\right)\right]\ ,\\
\Gamma^{I}_{2-2}&=&\Gamma^{I}_{22}\ ,
\end{eqnarray}
For $l=3$, it is straightforward to reach the following
\begin{eqnarray}
\Gamma^{I}_{30}&=&-\frac{\sqrt{7\pi}}{24}(1-\cos\xi)\left[(1+\cos\xi)(17+10\cos\xi+5\cos^2\xi)+48\log\left(\sin\frac{\xi}{2}\right)\right]\ ,\\
\Gamma^{I}_{31}&=&\frac{\sqrt{21\pi}}{48}\sin\xi(1-\cos\xi)\left[34+15\cos\xi+5\cos^2\xi+\frac{96}{1+\cos\xi}\log\left(\sin\frac{\xi}{2}\right)\right]\ ,\\
\Gamma^{I}_{3-1}&=&-\Gamma^{I}_{31}\ ,\\
\Gamma^{I}_{32}&=&-\frac{\sqrt{210\pi}}{48}(1-\cos\xi)\left[17-9\cos\xi-3\cos^2\xi-\cos^3\xi+48\left(\frac{1-\cos\xi}{1+\cos\xi}\right)\log\left(\sin\frac{\xi}{2}\right)\right]\ ,\\
\Gamma^{I}_{3-2}&=&\Gamma^{I}_{32}\ ,\\
\Gamma^{I}_{33}&=&\frac{\sqrt{35\pi}}{48}\frac{(1-\cos\xi)^2}{\sin\xi}\left[34-17\cos\xi-4\cos^2\xi-\cos^3\xi+96\left(\frac{1-\cos\xi}{1+\cos\xi}\right)\log\left(\sin\frac{\xi}{2}\right)\right]\ ,\\
\Gamma^{I}_{3-3}&=&-\Gamma^{I}_{33}\ .
\end{eqnarray}
These are plotted in Fig. \ref{GI}.
The generalized ORFs of total intensity are different from that of circular polarization in that the value for $m=0$ is non-trivial.
Then the $I$ mode ORFs for $(l,m)=(0,0),(1,0),(2,0)$ have value even in the case of $\xi=0$.
This implies that we can consider auto-correlation for a single pulsar.

\end{document}